\def\mycmd{2}

\if\mycmd1
\documentclass[11pt,onecolumn, draftcls]{IEEEtran}
\else 
\documentclass[lettersize,journal]{IEEEtran}
\fi

\usepackage[latin9]{inputenc}
\usepackage{tabularx}
\usepackage{threeparttable}

\usepackage{colortbl}    
\usepackage{pifont}     

\usepackage[british]{babel}
\usepackage{multicol}
\usepackage{array,ragged2e}
\usepackage{float}
\usepackage{mathtools}
\usepackage{amsmath}
\usepackage{amsthm}
\usepackage{amssymb}
\usepackage{graphicx}
\usepackage{wasysym}
\usepackage{setspace}
\usepackage{color}
\usepackage{bm}
\usepackage{cases}
\usepackage{makecell}
\usepackage{tikz}
\usepackage{flowchart}
\usepackage{amsfonts}
\usepackage{steinmetz}
\usetikzlibrary{matrix,shapes,arrows,positioning,chains}
\usepackage{lmodern,babel,adjustbox,booktabs,multirow}
\usepackage{makecell}
\usepackage{pbox}
\usepackage{epsfig}
\usepackage{subfigure}
\usepackage{tcolorbox}
\usepackage{enumitem}

\allowdisplaybreaks[4]

\makeatletter
\newcommand{\multiline}[1]{%
  \begin{tabularx}{\dimexpr\linewidth-\ALG@thistlm}[t]{@{}X@{}}
    #1
  \end{tabularx}
}

\floatstyle{ruled}
\newfloat{algorithm}{tbp}{loa}
\providecommand{\algorithmname}{Algorithm}
\floatname{algorithm}{\protect\algorithmname}

\theoremstyle{plain}

\theoremstyle{plain}

\theoremstyle{plain}

\usepackage{epsfig}
\usepackage[caption=false,font=normalsize,labelfont=sf,textfont=sf]{subfig}
\usepackage{cite}
\usepackage{stfloats}
\usepackage{graphicx}
\usepackage{multirow}
\usepackage{array}
\usepackage{graphicx}
\usepackage{epstopdf}
\usepackage{times}
\usepackage{algorithm}
\usepackage{algpseudocode}
\theoremstyle{remark}

\DeclareMathOperator*{\argmax}{arg\,max}
\makeatother

\algrenewcommand\algorithmicindent{1.0em}%
\providecommand{\lemmaname}{Lemma}
\providecommand{\propositionname}{Proposition}

\providecommand{\theoremname}{Theorem}
\providecommand{\theoremname}{Definition}
\newcommand{\rom}[1]{\uppercase\expandafter{\romannumeral #1\relax}}

\newcounter{problem}
\newcounter{save@equation}
\newcounter{save@problem}

\definecolor{lightergray}{gray}{0.9}
\definecolor{ForestGreen}{RGB}{34,139,34}  

\newcommand{\Xmark}{\textcolor{lightergray}{\ding{55}}}
\newcommand{\mycheck}{\textcolor{ForestGreen}{\ding{51}}}


\numberwithin{save@problem}{subsection}
\numberwithin{save@equation}{subsection}

\makeatletter
\renewcommand\paragraph[1]{%
    \vspace{.1cm}\noindent\textbf{#1.}
}
\makeatother

\begin{document}
\title{IMNet: Intercarrier Interference Mitigation Network for Integrated Sensing and Communication in Spectrally Efficient FDM Systems}
\author{Hyeonho Noh,~\IEEEmembership{Member,~IEEE}
\thanks{Hyeonho Noh is with the Department of Information and Communication Engineering, Hanbat National University, Republic of Korea (e-mail: hhnoh@hanbat.ac.kr). 
}
}

\maketitle
\begin{abstract}\label{abstract}
Spectrally efficient frequency-division multiplexing (SEFDM) is an attractive waveform to improve communication spectral efficiency by compressing the subcarrier spacing, yet its use for integrated sensing and communication (ISAC) poses a fundamental sensing challenge. Specifically, the intentional loss of subcarrier orthogonality generates SEFDM-induced intercarrier interference (S-ICI), which combines with Doppler-induced ICI (D-ICI) from moving targets to blur range--velocity maps and severely degrade sensing accuracy. Building on multi-user multi-input-multi-output (MIMO) SEFDM systems, this paper develops a model-driven ISAC framework that supports spectrally efficient multi-user communication while mitigating both S-ICI and D-ICI in sensing. To this end, an intercarrier interference mitigation network (IMNet) is proposed, which exploits the distinct physical structures of the two interferences. A bank of Doppler correction filters first compensates the velocity-dependent D-ICI over multiple Doppler hypotheses, and an axial-attention network subsequently suppresses the residual D-ICI and the long-range S-ICI to recover reliable sensing signals. To further improve range and velocity estimation accuracy, IMNet with local refinement (IMNet-LR) is proposed, which performs maximum-likelihood refinement with nuisance projection around the IMNet detections to achieve sub-cell precision without an exhaustive global search. Simulation results show that IMNet-LR achieves near-maximum-likelihood range and velocity estimation accuracy with more than three orders of magnitude lower execution time compared to conventional detection methods.
\end{abstract}

\begin{IEEEkeywords}
Integrated sensing and communication (ISAC), spectrally efficient FDM (SEFDM), intercarrier interference (ICI) mitigation network (IMNet), radar target detection
\end{IEEEkeywords}

\section{Introduction}
\label{sec:introduction}
Wireless networks are evolving from passive data-delivery systems into infrastructures that support intelligent agents operating in the physical world \cite{Bariah24_WCM,Xiao25_CM}. This evolution is being accelerated by physical artificial intelligence (AI), in which autonomous robots and other embodied agents must perceive, reason, and act in dynamic environments \cite{Yang25_VTM}. Their safe and coordinated operation requires not only reliable data exchange but also continuous localization, object tracking, motion estimation, and situational awareness \cite{Zhang22_CST}. Integrated sensing and communication (ISAC) supports these requirements by using a common waveform, spectrum, and transceiver platform for both data transmission and environmental sensing \cite{Liu22_JSAC,Liu20_TCOM_JRC}. ISAC can therefore transform wireless infrastructure into a distributed perception layer for physical AI rather than merely a medium for communication.

Among the waveforms available for practical ISAC, orthogonal frequency-division multiplexing (OFDM) has emerged as the most widely studied option because its two-dimensional time--frequency structure is inherently suited to joint communication and sensing \cite{Wei23_iotj}. The orthogonal time--frequency structure of OFDM enables efficient separation of target echoes in the range--Doppler domain, making it particularly amenable to joint range and velocity estimation \cite{strum11}. Moreover, the occupied bandwidth and coherent observation interval govern the attainable range and velocity resolutions, respectively, while subcarrier orthogonality enables efficient Fourier-domain processing \cite{Xiao24_TSP,Keskin23_TSP_PN,Xu23_TSP,Hu24_TWC}. Building on the same multicarrier principle, spectrally efficient frequency-division multiplexing (SEFDM) relaxes strict subcarrier orthogonality and increases the spectral overlap among adjacent subcarriers, allowing more data-bearing subcarriers to be accommodated within a given bandwidth. SEFDM thereby improves communication spectral efficiency while retaining the underlying time--frequency structure that supports joint range and Doppler estimation \cite{Mirabella26_TWC}. This combination of enhanced spectral efficiency and inherent sensing capability makes SEFDM a promising waveform for next-generation ISAC systems.

\begin{table}
\centering
    \caption{Comparison of the proposed scheme with recent works.}
    \vspace{-5pt}
    \begin{threeparttable}
    \adjustbox{width= \if 1\mycmd 0.6 \else 1.0 \fi \columnwidth}{
    \begin{tabular}{ccccccc}
    \toprule
    \textbf{Ref} & \textbf{Scheme} & \textbf{Sens.} & \textbf{Comm.} & \textbf{Compl.} & \textbf{ICI suppr.} & \textbf{SEFDM} \\ \midrule[\heavyrulewidth]\midrule[\heavyrulewidth]
    \arrayrulecolor{lightgray}
    \cite{Mirabella23_TCOM} & ML & \LEFTcircle & \LEFTcircle & \Circle & \Xmark & \Xmark \\ \cline{1-7}
    \cite{Zhang20} & ML & \CIRCLE & \LEFTcircle & \Circle & \mycheck & \Xmark \\ \cline{1-7}
    \cite{Zhang24_TWC, Keskin24_TWC_OTFS} & ML & \LEFTcircle & \LEFTcircle & \Circle & \Xmark & \Xmark \\ \cline{1-7}
    \cite{Xie21} & SS & \LEFTcircle & \LEFTcircle & \Circle & \Xmark & \Xmark \\ \cline{1-7}
    \cite{Liu20, Hu24_TWC} & SS & \LEFTcircle & \LEFTcircle & \LEFTcircle & \Xmark & \Xmark \\ \cline{1-7}
    \cite{strum11, Tian17, Sit18, Xiao24_TSP} & FFT & \Circle & \LEFTcircle & \CIRCLE & \Xmark & \Xmark \\ \cline{1-7}
    \cite{Xu23_TSP, Liyanaarachchi24_TWC, Duan24_TVT} & FFT & \LEFTcircle & \Circle & \CIRCLE & \Xmark & \Xmark \\ \cline{1-7}
    \cite{hakobyan18} & FFT & \LEFTcircle & \Circle & \CIRCLE & \mycheck & \Xmark \\ \cline{1-7}
    \cite{Keskin21} & ML, FFT & \CIRCLE & \LEFTcircle & \Circle & \mycheck & \Xmark \\ \cline{1-7}
    \cite{Noh23, Park24_TVT} & FFT & \LEFTcircle & \LEFTcircle & \LEFTcircle & \mycheck & \Xmark \\ \cline{1-7}
    \cite{Mirabella26_TWC} & ML & \CIRCLE & \CIRCLE & \Circle & \Xmark & \mycheck \\ \cline{1-7}
    \cite{DCFNet_TWC2026} & DL, FFT & \CIRCLE & \LEFTcircle & \LEFTcircle & \mycheck & \Xmark \\ \midrule
    \textbf{IMNet} & \textbf{DL, ML, FFT} & \CIRCLE & \CIRCLE & \LEFTcircle & \mycheck & \mycheck \\
    \arrayrulecolor{black}
    \bottomrule
    \end{tabular}
    }
    \vspace{0.02cm}
    \begin{flushleft}
    \footnotesize
    * SS: Subspace, Sens.: Sensing performance, \\
    Comm.: Communication performance, Compl.: Complexity. \\
    $\CIRCLE$, $\LEFTcircle$, and $\Circle$ denote high, moderate, and low performance, respectively.
    \end{flushleft}
    \end{threeparttable}
    \label{tab:compare}
\end{table}

\subsection{Related Works}
\label{subsec:related_works}

Various approaches have been proposed for estimating target range and velocity in multicarrier ISAC systems, as summarized in Table~\ref{tab:compare}. 

\paragraph{Multicarrier ISAC sensing}
Multicarrier sensing techniques have largely been developed around OFDM, whose time-frequency structure enables efficient joint range and Doppler estimation. The fast Fourier transform (FFT)-based ISAC receiver in \cite{strum11} decouples target delay and Doppler into phase progressions across subcarriers and symbols, allowing the range-Doppler map to be efficiently constructed through Fourier processing. This principle has been extended to multi-user (MU) multiple-input multiple-output (MIMO) systems for joint waveform and beamforming design \cite{Liu20_TSP_BF,Xu23_TSP} and interference management \cite{Sit18}. Despite its computational efficiency, FFT-based sensing is confined to discrete delay-Doppler bins determined by the finite numbers of subcarriers and symbols, thereby limiting estimation accuracy. To overcome this limitation, parametric and super-resolution methods have been developed, including MIMO range-velocity estimation \cite{Xiao24_TSP}, tensor-based estimation \cite{Zhang24_TWC}, estimation of signal parameters via rotational invariance techniques (ESPRIT)-type approaches \cite{Liu20,Hu24_TWC}, and maximum likelihood (ML)-based refinements \cite{Mirabella23_TCOM,Keskin24_TWC_OTFS}. Though these methods can provide sub-bin accuracy, their complexity or sensitivity to snapshot support and model mismatch can become problematic in dense scenes. The work in \cite{Noh26_PASS} further employed a physics-guided learning approach for target localization in wideband OFDM systems affected by hardware-induced dispersion, but its formulation is limited to single-symbol angle--range localization.

SEFDM has also been explored for ISAC, with \cite{Zhang24_SPAWC} experimentally demonstrating a MU MIMO-SEFDM system and its spectral-efficiency gains through over-the-air measurements. For SEFDM-based sensing, the loss of subcarrier orthogonality inherently introduces SEFDM-induced ICI (S-ICI), which has been addressed using model-based estimation and interference cancellation techniques \cite{Mirabella26_TWC,Mirabella26_SETFS}. Although these approaches account for the interference arising from non-orthogonal waveform compression, they do not explicitly address Doppler-induced ICI (D-ICI) caused by target motion. Moreover, their reliance on iterative ML-based refinement or spectral cancellation incurs substantial computational complexity.

\paragraph{ICI-robust ISAC} ICI-robust processing attempts to recover sensing performance when target motion or waveform mismatch breaks OFDM orthogonality. ICI-free OFDM radar processing can suppress D-ICI through dedicated symbol structures \cite{hakobyan18}, but this strategy significantly sacrifies communication data rate. Joint range and velocity estimation under intrapulse and intersubcarrier Doppler effects has also been analyzed in \cite{Zhang20}. Another direction exploits, rather than simply suppresses, ICI in MIMO-OFDM joint radar-communications \cite{Keskin21}. Nevertheless, both works \cite{Zhang20,Keskin21} depend on ML search, which leads to prohibitive computational complexity. Communication-compatible OFDM radar receivers with ICI mitigation have been further developed in \cite{Noh23,Park24_TVT}, and DCFNet \cite{DCFNet_TWC2026} suppresses D-ICI in MU MIMO-OFDM sensing through a Doppler correction filter (DCF) bank and a U-Net. Nevertheless, all of these receivers are tied to the OFDM model and are therefore not well suited to mitigating S-ICI.

\subsection{Challenges}
\label{subsec:challenges}
The above studies indicate two challenges that must be addressed jointly. First, SEFDM improves spectral efficiency by intentionally breaking subcarrier orthogonality; the resulting S-ICI raises the interference floor across the range-Doppler map, making weak target signals difficult to distinguish. Second, exhaustive ML and successive interference cancellation (SIC)-based refinements can approach high estimation accuracy, but their complexity increases with the search grid and number of targets, which makes dense low-latency sensing difficult. Motivated by these unresolved challenges, this study poses a key research question:

\begin{tcolorbox}[colframe=black, colback=white, height=1.4cm, boxrule=0.4mm, halign=center, valign=center]
\textit{\textbf{How can high-precision sensing be achieved with low computational complexity under SEFDM-induced ICI?}}
\end{tcolorbox}


\subsection{Contributions}
\label{subsec:contributions}

This paper develops a MU MIMO-SEFDM ISAC framework that exploits SEFDM for improved communication spectral efficiency while addressing the resulting S-ICI in radar sensing. To this end, IMNet, a model-driven learning receiver that combines the known SEFDM signal structure with learning-based interference mitigation, is proposed.

The contributions are summarized as follows:
\begin{itemize}[leftmargin=*]
\item A MU MIMO-SEFDM ISAC framework is developed in which transmit beamforming supports both communication and sensing. To maintain sufficient target sensing gain while maximizing communication spectral efficiency, a sum-rate maximization problem is formulated subject to power constraints and a minimum sensing beampattern gain toward the focal directions. The resulting beamformer provides a controllable communication--sensing operating point and strengthens the sensing signal before IMNet processing. Although the S-ICI couples the subcarriers, the problem is recast through the weighted minimum mean-square error (WMMSE) method and the precoder update is shown to be decoupled across subcarriers, yielding an efficient closed-form solution for each subcarrier.
\item IMNet, a model-driven detection architecture that exploits the distinct structures of D-ICI and S-ICI rather than treating them as generic interference, is proposed. IMNet first employs a DCF bank to generate multiple velocity-compensated range--velocity maps, such that targets affected by severe D-ICI can be recovered in maps associated with nearby Doppler hypotheses. These complementary observations are then jointly processed by a structure-aware ICI rejection head. In particular, IMNet selectively exploits the DCF outputs according to their residual interference levels, while range-wise attention mitigates the long-range S-ICI coupling across distant range cells and Doppler-wise attention suppresses the residual D-ICI after Doppler compensation. A detection head then transforms the ICI-mitigated representation into a target-confidence map, enabling robust target detection under severe S-ICI and D-ICI.
\item To improve the grid-limited estimation accuracy of IMNet without resorting to exhaustive likelihood search, IMNet with local refinement (IMNet-LR) is introduced. IMNet detections serve as initial target candidates, around which a local ML refinement with nuisance projection is performed. By restricting the likelihood evaluation to small neighborhoods of detected targets, IMNet-LR provides sub-cell range and velocity estimates while avoiding a fine-grid search over the entire range-Doppler domain.
\item Extensive simulations demonstrate that the proposed IMNet framework achieves substantial improvements in both detection accuracy and computational efficiency relative to conventional detectors. Its detection performance approaches the ideal interference-free bound, confirming its effectiveness in practical SEFDM ISAC scenarios. Furthermore, IMNet-LR achieves the range and velocity accuracy of an exhaustive likelihood search while running roughly three orders of magnitude faster.

\end{itemize}

\paragraph{Notations} $(\cdot)^{*}$, $(\cdot)^\text{T}$, $(\cdot)^\text{H}$, and $(\cdot)^\dagger$ are complex
conjugate, transpose, conjugate transpose, and Moore--Penrose pseudo-inverse, respectively.
For a matrix $\mathbf{A}$, $\mathbf{A}^{-1}$, $\mathrm{Tr}(\mathbf{A})$, and $\left[\mathbf{A}\right]_{(i,j)}$ are the inverse, trace, and $(i,j)$-th entry of $\mathbf{A}$, respectively, and $\mathbf{A}\succeq\mathbf{0}$ and $\mathbf{A}\succ\mathbf{0}$ indicate that $\mathbf{A}$ is positive semidefinite and positive definite. The symbol $./$ is the elementwise complex division and $*$ is the two-dimensional convolution. The $N \times N$ identity matrix is denoted by $\mathbf{I}_N$, and $\mathbf{1}$ is the all-ones vector. The diagonal matrix with diagonal elements $(a_1, a_2, \ldots,a_m)$ is denoted by $\text{diag}(a_1, a_2, \ldots, a_m)$.
$\mathcal{CN} (\mu, \Sigma)$ is the complex normal distribution with mean $\mu$ and covariance matrix $\Sigma$. $\mathbb{R}^{n \times m}$ and $\mathbb{C}^{n \times m}$ are the $(n \times m)$-dimensional real and complex spaces, respectively. $\mathrm{Re}\{\cdot\}$ and $\mathrm{Im}\{\cdot\}$ are the real and imaginary parts, $\left\| \cdot \right\|_2$ is the Euclidean norm, $\lfloor \cdot \rfloor$ is the floor operator.

\section{Scenario and Protocol}
\label{sec:system_model}

\begin{figure*}[t]
\centering
\includegraphics[width=\textwidth]{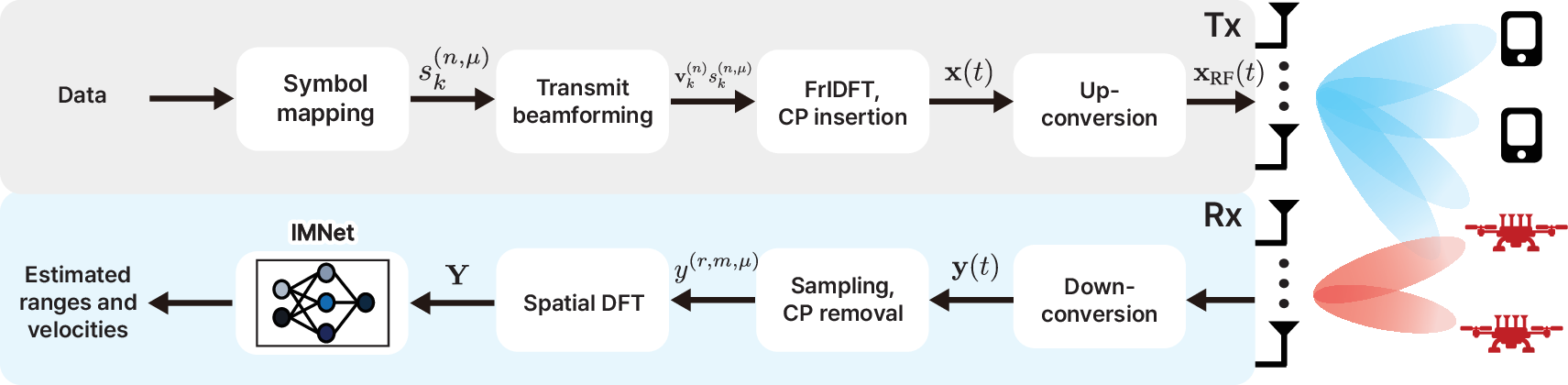}
\caption{Architecture of the considered MU-MIMO-SEFDM ISAC system.}
\label{fig:signal_process}
\end{figure*}

The proposed MU MIMO-SEFDM ISAC system, consisting of a base station (BS), $J$ radar targets, and $K$ communication users, is considered. The BS is equipped with a uniform linear array (ULA) comprising $N_\text{T}$ transmit antennas and $N_\text{R}$ receive antennas, while each user is equipped with a single receive antenna.
Compared to OFDM, SEFDM compresses the subcarrier spacing from $\Delta f$ to $\beta\Delta f$ with bandwidth compression factor $0 < \beta \leq 1$, thereby accommodating more subcarriers within a given bandwidth at the cost of the loss of subcarrier orthogonality and the resulting inherent ICI. The BS simultaneously serves the users and performs target sensing. Its primary objective is to estimate the ranges and velocities of targets while transmitting data to $K$ users. The ISAC system operates in two stages:

\paragraph{ISAC signal transmission} The BS transmits SEFDM signals embedded with communication symbols, which are used both for downlink communication and target sensing within predefined focal angles. To improve the downlink data rate and enhance beamforming gain toward focal angles, the BS applies transmit beamforming.

\paragraph{Sensing signal reception} The BS retains the full per-antenna time-domain samples across its $N_\text{R}$ receive antennas and sweeps its receive beam over the angular domain; for each observed angle of arrival (AoA), it forms a two-dimensional range-velocity map through 2D fractional Fourier processing.

Far-field point targets with constant radial velocities are considered, and the BS-to-user channels include multipath components from scatterers in the propagation environment.

\subsection{System and Signal Model}
\label{subsec:system_model}
Over $N_\text{sym}$ consecutive SEFDM symbols, the baseband transmit signal vector is given by
\begin{align} \label{def:SEFDM_signal}
     \mathbf{x}(t) &= \frac{1}{\sqrt{N_\text{c}}} \sum_{\mu=0}^{N_\text{sym}-1}\sum_{n\in\mathcal{N}_\text{c}}\sum_{k\in\mathcal{K}} \mathbf{v}_{k}^{(n)} s_{k}^{(n,\mu)} \nonumber \\
     & \hspace{30pt} \times {e}^{j2\pi n \beta \Delta f t_\mu}\, \text{rect}\left( \frac{t_\mu+T_{\text{CP}}}{T_{\text{OFDM}}} \right),
\end{align}
where $N_\text{c}$ is the number of SEFDM subcarriers; $\mathcal{K} = \{0,\ldots,K-1\}$; $\mathbf{v}_{k}^{(n)} \in \mathbb{C}^{N_\text{T} \times 1}$ is the transmit beamformer for user $k$ on subcarrier $n$; $s_{k}^{(n,\mu)}$ is the communication symbol of user $k$ on subcarrier $n$ in the $\mu$-th symbol; $\Delta f = 1/T$ is the OFDM-equivalent subcarrier spacing serving as the reference grid; $T$ is the SEFDM signal duration; $T_\text{CP}$ is the cyclic prefix (CP) duration; $T_\text{OFDM} = T + T_\text{CP}$; $t_\mu = t - \mu T_\text{OFDM}$; and $\text{rect}(\cdot)$ is the rectangular function defined as
\begin{align}
    \text{rect} \Big( \frac{t}{T} \Big) =\begin{cases}
    1, & \text{for} ~ 0 \leq t < T,\\
    0, & \text{otherwise}.
  \end{cases}
\end{align}
The channel coherence time is assumed to exceed $N_\text{sym} T_\text{OFDM}$, so that $\mathbf{v}_k^{(n)}$ is constant over the transmission. The BS transmits the up-converted signal $\mathbf{x}_\text{RF}(t) = \mathbf{x}(t) e^{j 2 \pi f_\text{c} t}$.

\subsection{Communication SINR}
\label{subsec:communication_SINR}
Each user down-converts the received radio-frequency (RF) signal to baseband and applies standard carrier frequency recovery. It is assumed that the carrier frequency offset (CFO) is compensated by conventional synchronization \cite{Schmidl97_TCOM, Moose94_TCOM} so that its residual impact on the signal-to-interference-plus-noise ratio (SINR) is negligible. After CP removal, each user applies the fractional discrete Fourier transform (FrDFT) of parameter $\beta$ to project the time-domain samples onto the SEFDM subcarrier grid. For $\beta<1$, the SEFDM subcarriers are non-orthogonal, and the FrDFT output associated with subcarrier $n$ contains contributions from the other transmitted subcarriers. The coupling from subcarrier $m$ to subcarrier $n$, for $m,n \in \mathcal{N}_\text{c}$, is characterized by
\begin{align}\label{def:SEFDM_ICI_coeff}
    \rho_{m,n}(\beta) = \frac{1}{N_\text{c}} \sum_{\ell=0}^{N_\text{c}-1} e^{j 2\pi (m-n)\beta\ell / N_\text{c}}.
\end{align}
For each $k \in \mathcal{K}$, the SINR of user $k$ on subcarrier $n$ reads
\begin{align}\label{eq:sinr_def}
    \gamma_k^{(n)} &= \frac{ \Big\|(\mathbf{h}_k^{(n)})^\text{H} \mathbf{v}_k^{(n)}\Big\|^2 }{ I_k^{(n)} + J_k^{(n)}(\beta) + \sigma_k^2},
\end{align}
where
\begin{align}
    I_k^{(n)} &= \sum_{\ell\in\mathcal{K} \backslash \{ k\}} \Big\|(\mathbf{h}_k^{(n)})^\text{H} \mathbf{v}_\ell^{(n)}\Big\|^2,\\
    J_k^{(n)}(\beta) &= \sum_{m\in\mathcal{N}_\text{c}\backslash\{n\}} |\rho_{m,n}(\beta)|^2 \sum_{\ell\in\mathcal{K}} \Big\|(\mathbf{h}_k^{(m)})^\text{H} \mathbf{v}_\ell^{(m)}\Big\|^2, \label{def:J_kn}
\end{align}
$\mathbf{h}_k^{(n)} \in \mathbb{C}^{N_\text{T} \times 1}$ is the BS-to-user frequency-domain channel vector evaluated at subcarrier $n$, and $\sigma_k^2$ is the noise variance. The term $I_k^{(n)}$ captures the multi-user interference within subcarrier $n$, and $J_k^{(n)}(\beta)$ collects the S-ICI leaking from the other subcarriers.

\subsection{Radar Received Signal}
\label{subsec:radar_received_signal}
The reflected waveform is collected by the $N_\text{R}$-element receive ULA. The standard far-field, monostatic point-target model with constant path gains, negligible range migration, and a sufficient CP is used \cite{Xiao24_TSP,hakobyan18,Noh23,Keskin21}. Thus the $i$-th path has delay $\tau_i=2R_i/c_0$, Doppler $f_{\mathrm{D},i}=-2v_i f_\text{c}/c_0$, and array response $\mathbf{A}_i=\mathbf{a}_\text{R}(\theta_i)\mathbf{a}_\text{T}^{\mathrm H}(\theta_i)$ with $[\mathbf{a}_\text{M}(\theta)]_r=e^{-j2\pi r\Delta\sin\theta}$ for $\text{M}\in\{\text{T},\text{R}\}$.

Following down-conversion, the baseband reflected signal can be represented by
\begin{align}\label{def:down_converted_signal}
    \mathbf{y}(t)
    &= \sum_{i\in\mathcal{P}} a_i e^{-j2\pi f_\text{c}\tau_i} e^{j2\pi f_{\mathrm{D},i}t}
    \mathbf{A}_i\mathbf{x}(t-\tau_i) + \mathbf{z}(t).
\end{align}
After CP removal and sampling at $t_\mu=mT/N_\text{c}$, the $r$-th receive-antenna sample is given by
\begin{align}\label{extended_disc_radar}
    y^{(r,m,\mu)} &= \sum_{i\in\mathcal{P}} \sum_{n\in\mathcal{N}_\text{c}}\sum_{k\in\mathcal{K}} \frac{\bar{a}_i}{\sqrt{N_\text{c}}}\, [\mathbf{a}_\text{R}(\theta_i)]_r\, \mathbf{a}_\text{T}^\text{H}(\theta_i)\mathbf{v}_k^{(n)} s_k^{(n,\mu)} \nonumber \\
    &\times e^{j2\pi n\beta m/N_\text{c}} e^{-j2\pi n\beta\bar{\tau}_i} \nonumber \\
    &\times e^{j2\pi\bar{f}_{\text{D},i} m/N_\text{c}} e^{j2\pi\bar{f}_{\text{D},i}\mu\alpha} + \hat{z}^{(r,m,\mu)},
\end{align}
where $\bar a_i=a_i e^{-j2\pi f_\text{c}\tau_i}$, $\bar\tau_i=\tau_i/T$, $\bar f_{\mathrm{D},i}=f_{\mathrm{D},i}T$, $\alpha=T_\text{OFDM}/T$, and $\hat z^{(r,m,\mu)}\sim\mathcal{CN}(0,\sigma^2)$. 
Stacking \eqref{extended_disc_radar} over $m$ and $\mu$ gives $\mathbf{Y}_r\in\mathbb{C}^{N_\text{c}\times N_\text{sym}}$ with
\begin{align}\label{disc_received_signal_mat_form}
    \mathbf{Y}_r
    &= \sum_{i\in\mathcal{P}} \bar{a}_i [\mathbf{a}_\text{R}(\theta_i)]_r\,
    \mathbf{D}_{\text{I}}\big(\bar{f}_{\text{D},i}\big) \mathbf{F}_{N_\text{c},\beta}^\text{H}
    \mathbf{D}_{\text{R}}^*\big(\beta\bar{\tau}_{i}\big) \tilde{\mathbf{S}}_i
    \mathbf{D}_{\text{v}}\big(\bar{f}_{\text{D},i}\big) \nonumber\\
    &\hspace{180pt}+ \mathbf{Z}_r,
\end{align}
where
\begin{align}
    [\tilde{\mathbf{S}}_i]_{(p,q)} & = \mathbf{a}_\text{T}^\text{H}(\theta_i)\sum\limits_{k\in\mathcal{K}} \mathbf{v}_k^{(p)} s_k^{(p,q)} \label{subeq:s_def}, \\
    \mathbf{D}_{\text{I}}(f) & = \text{diag} \big( 1, e^{j2\pi \frac{f}{N_\text{c}}}, \ldots, e^{j2\pi \frac{f}{N_\text{c}} (N_\text{c}-1)} \big), \\
    \mathbf{D}_{\text{R}}(\tau) & = \text{diag} \big( 1, e^{j2\pi \tau}, \ldots, e^{j2\pi \tau (N_\text{c}-1)} \big), \\
    \mathbf{D}_{\text{v}}(f) & = \text{diag} \big( 1, e^{j2\pi f \alpha}, \ldots, e^{j2\pi f \alpha (N_\text{sym}-1)} \big),
\end{align}
and the $N$-point FrDFT matrix $\mathbf{F}_{N,\beta}$ of parameter $\beta$ has entries, for $0 \le p, q \le N-1$,
\begin{align}\label{def:FrDFT_matrix}
    \left[\mathbf{F}_{N,\beta}\right]_{(p,q)} = \frac{1}{\sqrt{N}} e^{-j \frac{2 \pi \beta}{N}pq},
\end{align}
and $[\mathbf{Z}_r]_{(m,\mu)}\sim\mathcal{CN}(0,\sigma^2)$. 
Equation \eqref{disc_received_signal_mat_form} exposes the two frequency-domain distortions that motivate IMNet: D-ICI from $\mathbf{D}_{\text{I}}(\bar f_{\mathrm D,i})$ and S-ICI from the non-orthogonal fractional inverse discrete Fourier transform (FrIDFT) $\mathbf{F}_{N_\text{c},\beta}^{\mathrm H}$.

\subsection{Joint Angle-Range-Velocity Processing}
\label{subsec:2D FFT process for target detection}

The per-antenna model in \eqref{disc_received_signal_mat_form} preserves angle, fast-time, and slow-time dimensions. The BS sweeps its receive beam across the angular domain by a spatial DFT and, for each observed AoA, forms a two-dimensional range-velocity map.

\paragraph{Step 1: Receive beamforming toward the observed AoA}
The BS steers a receive beam to the observed AoA $\theta$ by applying the conjugate array response $\mathbf{a}_\text{R}^*(\theta)$ along the receive-antenna axis. For each $(m,\mu)$, the beamformed sample is
\begin{align}\label{eq:spatial_dft}
    \mathring{y}^{(m,\mu)} = \frac{1}{\sqrt{N_\text{R}}}\sum_{r=0}^{N_\text{R}-1} y^{(r,m,\mu)}\, e^{\,j 2\pi r \Delta \sin\theta}.
\end{align}
Sweeping $\theta$ over the angular domain scans the scene, and the remaining analysis is carried out for one such observation angle. The angular Dirichlet response of the steered beam to a path arriving from $\theta_i$ is
\begin{align}\label{eq:kappa}
    \kappa(\theta_i) = \frac{1}{\sqrt{N_\text{R}}}\sum_{r=0}^{N_\text{R}-1} e^{-j 2\pi r \Delta (\sin\theta_i - \sin\theta)},
\end{align}
which peaks at $\theta_i = \theta$ and decays for off-boresight paths. Collecting the beamformed data into $\mathbf{Y}\in\mathbb{C}^{N_\text{c}\times N_\text{sym}}$ yields
\begin{align}\label{eq:Y_na}
    \mathbf{Y}
    \hspace{-3pt}=\hspace{-3pt} \sum_{i\in\mathcal{P}} \bar a_i \kappa(\theta_i)\,
    \mathbf{D}_\text{I}(\bar f_{\text{D},i})\mathbf{F}_{N_\text{c},\beta}^\text{H}
    \mathbf{D}_\text{R}^*(\beta\bar\tau_i) \tilde{\mathbf{S}}_i
    \mathbf{D}_\text{v}(\bar f_{\text{D},i})
    \hspace{-3pt}+\hspace{-3pt} \mathbf{Z},
\end{align}
where $\mathcal{P}$ is the multipath set of \eqref{disc_received_signal_mat_form} and the angular weight $\kappa(\theta_i)$ concentrates the beam response on paths aligned with $\theta$ while suppressing off-boresight paths.

\paragraph{Step 2: ICI decomposition at the observed AoA}
For paths within the beam mainlobe around $\theta$, it holds  $\tilde{\mathbf{S}}_i\approx\tilde{\mathbf{S}}$ with $[\tilde{\mathbf{S}}]_{(p,q)} = \mathbf{a}_\text{T}^\text{H}(\theta)\sum_{k\in\mathcal{K}}\mathbf{v}_k^{(p)} s_k^{(p,q)}$. By defining the deviation term  $\Delta\mathbf{F}_\beta^\text{H} = \mathbf{F}_{N_\text{c},\beta}^\text{H} - \mathbf{F}_{N_\text{c},1}^\text{H}$, \eqref{eq:Y_na} becomes
\begin{align}\label{disc_received_signal_mat_form_divide}
    \mathbf{Y} 
    &= \sum_{i\in\mathcal{P}} \bar a_i \kappa(\theta_i)\, \mathbf{F}_{N_\text{c}}^\text{H} \mathbf{D}_\text{R}^*(\beta\bar\tau_i)\tilde{\mathbf{S}}\mathbf{D}_\text{v}(\bar f_{\text{D},i}) \nonumber \\ 
    &+ \mathbf{Y}^\text{leak} + \sum_{i\in\mathcal{P}}\big(\mathbf{Y}_{\text{D-ICI},i} + \mathbf{Y}_{\text{S-ICI},i}\big) + \mathbf{Z},
\end{align}
where
\begin{align}
    \mathbf{Y}_{\text{D-ICI},i}
    &= \bar a_i \kappa(\theta_i)\,(\mathbf{D}_\text{I}(\bar f_{\text{D},i})-\mathbf{I}_{N_\text{c}})
    \mathbf{F}_{N_\text{c}}^\text{H}\mathbf{D}_\text{R}^*(\beta\bar\tau_i)
    \tilde{\mathbf{S}}\mathbf{D}_\text{v}(\bar f_{\text{D},i}), \\
    \mathbf{Y}_{\text{S-ICI},i}
    &= \bar a_i \kappa(\theta_i)\,\mathbf{D}_\text{I}(\bar f_{\text{D},i})
    \Delta\mathbf{F}_\beta^\text{H}\mathbf{D}_\text{R}^*(\beta\bar\tau_i)
    \tilde{\mathbf{S}}\mathbf{D}_\text{v}(\bar f_{\text{D},i}).
\end{align}
$\mathbf{Y}_{\text{D-ICI},i}$ and $\mathbf{Y}_{\text{S-ICI},i}$ are the D-ICI and S-ICI contributions of path $i$, respectively, and $\mathbf{Y}^\text{leak}$ collects the residual off-boresight (cross-angular) leakage left by the mainlobe approximation $\tilde{\mathbf{S}}_i\approx\tilde{\mathbf{S}}$. 

\paragraph{Step 3: range--Doppler processing} At the observed AoA $\theta$, the BS first applies the FrDFT along the fast-time dimension to obtain the SEFDM subcarrier-domain observations. It then removes the known data modulation through elementwise division by the effective transmit-symbol matrix $\tilde{\mathbf{S}}$. Finally, the BS applies the inverse FrDFT along the range dimension and the DFT along the slow-time dimension to form the range--Doppler image:
\begin{align}\label{eq:radar_image}
    \Bar{\mathbf{Y}} = \mathbf{F}_{N_\text{c},\beta}^\text{H}\big(\mathbf{F}_{N_\text{c},\beta}\mathbf{Y} ./ \tilde{\mathbf{S}}\big)\mathbf{F}_{N_\text{sym}}.
\end{align}
With the usual within-bin approximation $\tilde{\mathbf{S}}_i ./ \tilde{\mathbf{S}}\approx \mathbf{1} \times \mathbf{1}^{\mathrm T}$, \eqref{eq:radar_image} yields
\begin{align}\label{eq:radar_image_approx}
    \Bar{\mathbf{Y}}
    &\approx \sum_{i\in\mathcal{P}}\bar a_i \kappa(\theta_i)
    \mathbf{e}^{*}_\text{R}(\beta\bar\tau_i)\mathbf{e}_\text{v}^{\mathrm T}(\bar f_{\text{D},i})
    + \Bar{\mathbf{Y}}^\text{leak} \nonumber\\
    &\quad + \sum_{i\in\mathcal{P}}
    \big(\Bar{\mathbf{Y}}_{\text{D-ICI},i}
    + \Bar{\mathbf{Y}}_{\text{S-ICI},i}\big)
    + \Bar{\mathbf{Z}},
\end{align}
where the DFT response vectors are defined by
\begin{align}
    [\mathbf{e}_\text{R}(\bar\tau)]_m
    &= \frac{1}{\sqrt{N_\text{c}}}\sum_{n=0}^{N_\text{c}-1}e^{j2\pi(\bar\tau-m/N_\text{c})n},\\
    [\mathbf{e}_\text{v}(\bar f)]_\mu
    &= \frac{1}{\sqrt{N_\text{sym}}}\sum_{q=0}^{N_\text{sym}-1}e^{j2\pi(\bar f\alpha-\mu/N_\text{sym})q}.
\end{align}
Thus the desired path produces a range-Doppler peak, while two structured impairments remain as sidelobes: the D-ICI $\Bar{\mathbf{Y}}_{\text{D-ICI},i}$ and the S-ICI $\Bar{\mathbf{Y}}_{\text{S-ICI},i}$, which IMNet corrects in Sec.~\ref{sec:proposed_scheme}. 

\section{Proposed Scheme: IMNet}
\label{sec:proposed_scheme}

\begin{figure*}[t]
\centering
\includegraphics[width=\textwidth]{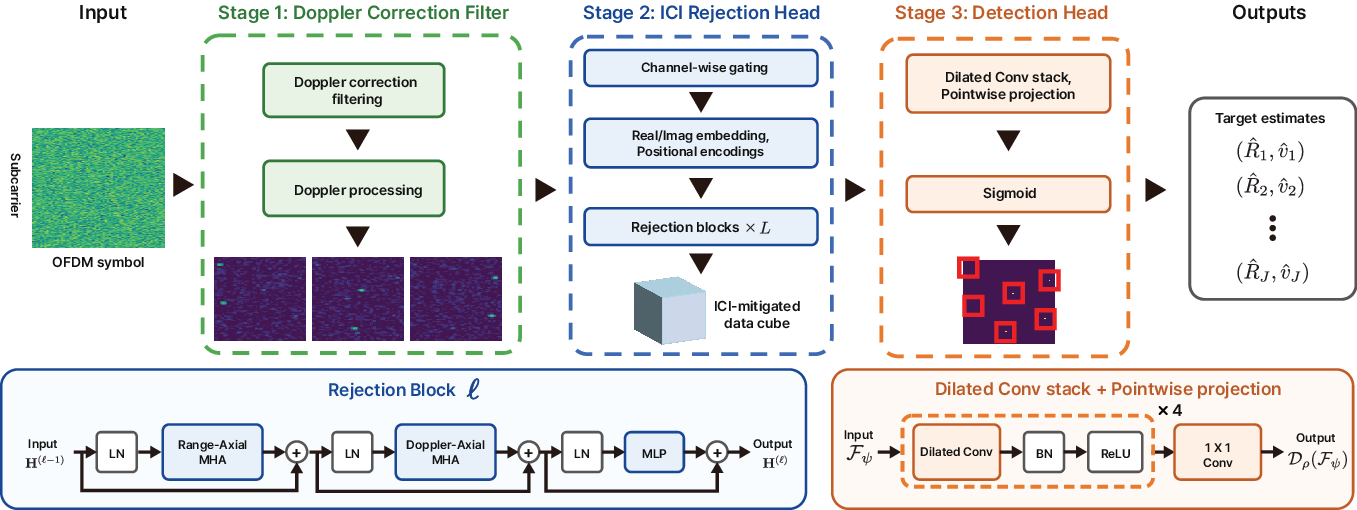}
\caption{Architecture of IMNet, consisting of three stages: (i) a DCF bank that compensates Doppler-induced ICI over multiple velocity hypotheses, (ii) an ICI rejection head that fuses the DCF outputs and suppresses residual S-ICI and D-ICI through channel-wise gating and axial self-attention, and (iii) a detection head that produces a target-confidence map over the range-Doppler grid.}
\label{fig:network}
\end{figure*}

This section presents IMNet, a model-driven deep architecture that suppresses the impairments exposed in Sec.~\ref{subsec:2D FFT process for target detection}: the D-ICI and the S-ICI. IMNet assigns each impairment to a dedicated stage tailored to its structure, and the three stages are cascaded as illustrated in Fig.~\ref{fig:network}.

\subsection{Stage 1: Doppler Correction Filter Bank}
\label{subsec:imnet_dcf}

Low-velocity targets induce negligible Doppler-ICI and are already visible in the range-Doppler map, whereas high-velocity targets generate strong ICI that raises the sidelobe floor and buries nearby targets. A DCF $\mathbf{D}_\text{I}^*(\bar f_{\text{D},k})$ shifts the low-ICI region from zero velocity to a hypothesized velocity $\bar f_{\text{D},k}$, so that fast targets around $\bar f_{\text{D},k}$ re-emerge from the floor, at the cost of contaminating the other velocity bands. Stage 1 therefore forms a bank of $N_v$ maps, each clean over a different velocity band, which Stage 2 fuses so that every target is recovered in at least one map.

\paragraph{Doppler correction filtering} Let $\mathcal{V} = \{\bar f_{\text{D},k}\}_{k\in\mathcal{N}_v}$ denote a velocity hypothesis grid of size $N_v$, with $\mathcal{N}_v = \{-\lfloor N_v/2\rfloor, \ldots, \lfloor N_v/2\rfloor - 1\}$. For each $k \in \mathcal{N}_v$, the BS derotates the fast-time samples by the hypothesized Doppler,
\begin{align}\label{eq:dcf}
    \mathbf{Y}^{(k)} = \mathbf{D}_\text{I}^*(\bar f_{\text{D},k})\,\mathbf{Y}.
\end{align}
Substituting \eqref{eq:Y_na} and using $\mathbf{D}_\text{I}^*(\bar f_{\text{D},k})\mathbf{D}_\text{I}(\bar f_{\text{D},i}) = \mathbf{D}_\text{I}(\bar f_{\text{D},i}-\bar f_{\text{D},k})$ yields
\begin{align}\label{eq:dcf_decompose}
    \mathbf{Y}^{(k)} & = \sum_{i\in\mathcal{P}} \bar a_i \kappa(\theta_i)\, \mathbf{D}_\text{I}(\bar f_{\text{D},i}-\bar f_{\text{D},k}) \mathbf{F}_{N_\text{c},\beta}^\text{H} \mathbf{D}_\text{R}^*(\beta\bar\tau_i) \tilde{\mathbf{S}} \nonumber \\
    & \hspace{15pt} \times \mathbf{D}_\text{v}(\bar f_{\text{D},i}) + \mathbf{D}_\text{I}^*(\bar f_{\text{D},k})\mathbf{Y}^\text{leak} + \mathbf{D}_\text{I}^*(\bar f_{\text{D},k})\mathbf{Z},
\end{align}
which recasts each target's Doppler-ICI in terms of its offset $\bar f_{\text{D},i}-\bar f_{\text{D},k}$ from the filter velocity. A target inside the compensated band ($\bar f_{\text{D},i} \approx \bar f_{\text{D},k}$) has $\mathbf{D}_\text{I}(\bar f_{\text{D},i}-\bar f_{\text{D},k}) \approx \mathbf{I}_{N_\text{c}}$, so its ICI vanishes and it re-emerges from the floor; targets outside the band keep the ICI term and stay buried in this map, but are recovered in the map whose band matches their velocity.

\paragraph{Doppler processing} The per-hypothesis radar image is then formed in analogy with \eqref{eq:radar_image}:
\begin{align}\label{eq:radar_image_k}
    \Bar{\mathbf{Y}}^{(k)} = \mathbf{F}_{N_\text{c},\beta}^\text{H}\big(\mathbf{F}_{N_\text{c},\beta}\,\mathbf{Y}^{(k)} ./ \tilde{\mathbf{S}}\big)\mathbf{F}_{N_\text{sym}}.
\end{align}
Stacking $\{\Bar{\mathbf{Y}}^{(k)}\}$ across the velocity hypotheses $\mathcal{N}_v$ produces the DCF-bank tensor $\Bar{\boldsymbol{\mathcal{Y}}}^{\text{DCF}} \in \mathbb{C}^{N_v \times N_\text{c} \times N_\text{sym}}$, i.e., the 2D range-velocity observation augmented with $N_v$ velocity-compensated channels. Since $\mathbf{D}_\text{I}^*(\bar f_{\text{D},k})$ acts only on $\mathbf{D}_\text{I}(\bar{f}_{\text{D},i})$, the DCF changes the ICI level of a target but not its position on the range-Doppler grid. All $N_v$ maps therefore share a common grid on which every target exists at the same cell. 

\subsection{Stage 2: ICI Rejection Head}
\label{subsec:imnet_deconv}

Stage 2 turns the $N_v$-channel DCF-bank tensor 
$\Bar{\boldsymbol{\mathcal{Y}}}^{\mathrm{DCF}}$ into a single coherent range-Doppler representation. Its design follows from an asymmetry between the two interference mechanisms. While the D-ICI can be largely compensated by the velocity-matched DCF bank in Stage~1, the SEFDM transform is non-unitary for $\beta<1$, i.e., 
$\mathbf{F}_{N_\text{c},\beta}^{H}\mathbf{F}_{N_\text{c},\beta} 
\neq \mathbf{I}_{N_\text{c}}$, so the conventional FrDFT/FrIDFT processing in \eqref{eq:radar_image_k} cannot fully remove the S-ICI. Stage~2 therefore applies axial attention along both range and Doppler: range-wise attention captures and suppresses the long-range S-ICI induced by SEFDM compression, while Doppler-wise attention further refines the residual Doppler-domain interference after DCF compensation.

Since the residual interference level varies across the $N_v$ DCF-bank
slices, Stage~2 first learns their relative contributions through a
channel-wise gating mechanism. For $i=0,\ldots,N_v-1$, the descriptor
$\mathbf{g}\in\mathbb{R}^{N_v}$ and the gate
$\mathbf{w}\in(0,1)^{N_v}$ are given by
\begin{align}
    [\mathbf{g}]_i
    &= \frac{1}{N_\text{c}N_\text{sym}}
    \sum_{m=0}^{N_\text{c}-1}
    \sum_{\mu=0}^{N_\text{sym}-1}
    \big|[\Bar{\mathbf{Y}}^{(i)}]_{m,\mu}\big|,
    \label{eq:se_desc}\\
    \mathbf{w}
    &= \sigma\big(
    \mathbf{W}_2\,\mathrm{ReLU}(\mathbf{W}_1\mathbf{g})
    \big),
    \label{eq:se_gate}
\end{align}
where $\sigma$ is the elementwise logistic function,
$\mathbf{W}_1\in\mathbb{R}^{d_\text{s}\times N_v}$,
$\mathbf{W}_2\in\mathbb{R}^{N_v\times d_\text{s}}$, and $d_\text{s}$
is the bottleneck width. The same gate is applied to the real and imaginary
components of each DCF slice, preserving their relative phase information.
The gated slices are then stacked into the real-valued tensor
$\boldsymbol{\mathcal{G}}
\in\mathbb{R}^{2N_v\times N_\text{c}\times N_\text{sym}}$ as
\begin{align}
    [\boldsymbol{\mathcal{G}}]_{i,m,\mu}
    &= [\mathbf{w}]_i\,
    \mathrm{Re}\big\{[\Bar{\mathbf{Y}}^{(i)}]_{m,\mu}\big\},
    \label{eq:stack_re}\\
    [\boldsymbol{\mathcal{G}}]_{N_v+i,m,\mu}
    &= [\mathbf{w}]_i\,
    \mathrm{Im}\big\{[\Bar{\mathbf{Y}}^{(i)}]_{m,\mu}\big\}.
    \label{eq:stack_im}
\end{align}

A $3\times3$ convolution with kernel
$\boldsymbol{\mathcal{W}}_\text{e}
\in\mathbb{R}^{d\times 2N_v\times3\times3}$
first extracts local range-Doppler features from
$\boldsymbol{\mathcal{G}}$ and embeds them into $d$ channels. Learned positional encodings $\mathbf{u}_\text{R}\in\mathbb{R}^{d\times N_\text{c}}$ and $\mathbf{u}_\text{D}\in\mathbb{R}^{d\times N_\text{sym}}$ are then added along the range and Doppler axes, respectively. For a channel index $c\in\{0,\ldots,d-1\}$, a range index $m\in\{0,\ldots,N_\text{c}-1\}$, and a Doppler index $\mu\in\{0,\ldots,N_\text{sym}-1\}$, the embedding $\mathbf{H}^{(0)} \in\mathbb{R}^{d\times N_\text{c}\times N_\text{sym}}$ is given by
\begin{align}\label{eq:embedding}
    [\mathbf{H}^{(0)}]_{c,m,\mu}
    =
    \big[
    \phi\big(
    \boldsymbol{\mathcal{W}}_\text{e}
    * \boldsymbol{\mathcal{G}}
    \big)
    \big]_{c,m,\mu}
    + [\mathbf{u}_\text{R}]_{c,m}
    + [\mathbf{u}_\text{D}]_{c,\mu},
\end{align}
where $*$ denotes 2D convolution over the range-Doppler grid with zero padding, and $\phi$ denotes a rectified linear unit (ReLU) activation followed by batch normalization. 

The tensor $\mathbf{H}^{(0)}$ is refined by $L$ rejection blocks, each
consisting of successive range- and Doppler-axis self-attention followed by
a pointwise multilayer perceptron (MLP). At the $\ell$-th block, the input is
$\mathbf{H}^{(\ell-1)}
\in\mathbb{R}^{d\times N_\text{c}\times N_\text{sym}}$.
For each Doppler index $\mu$, the range slice
$\mathbf{H}^{(\ell-1)}_{\text{R},\mu}
\in\mathbb{R}^{N_\text{c}\times d}$
treats the $N_\text{c}$ range cells as tokens, whereas for each range index
$m$, the Doppler slice
$\mathbf{H}^{(\ell-1)}_{\text{D},m}
\in\mathbb{R}^{N_\text{sym}\times d}$
treats the $N_\text{sym}$ Doppler cells as tokens.

For a unified description, let $a\in\{\mathrm{R},\mathrm{D}\}$ denote the
attention axis and let $s$ index the slices along the complementary axis.
The attention along each axis uses $n_\text{h}$ heads in parallel. For the
$h$th head of the $\ell$-th block, the query, key, and value matrices for
slice $s$ are constructed as
\begin{align}
    \mathbf{Q}^{(\ell)}_{a,s,h}
    &=
    \mathrm{LN}\left(\mathbf{H}^{(\ell-1)}_{a,s}\right)
    \mathbf{W}^{(\ell)}_{\mathrm{q},a,h},
    \label{eq:query}\\
    \mathbf{K}^{(\ell)}_{a,s,h}
    &=
    \mathrm{LN}\left(\mathbf{H}^{(\ell-1)}_{a,s}\right)
    \mathbf{W}^{(\ell)}_{\mathrm{k},a,h},
    \label{eq:key}\\
    \mathbf{V}^{(\ell)}_{a,s,h}
    &=
    \mathrm{LN}\left(\mathbf{H}^{(\ell-1)}_{a,s}\right)
    \mathbf{W}^{(\ell)}_{\mathrm{v},a,h},
    \label{eq:value}
\end{align}
where
$\mathbf{W}^{(\ell)}_{\mathrm{q},a,h},
 \mathbf{W}^{(\ell)}_{\mathrm{k},a,h},
 \mathbf{W}^{(\ell)}_{\mathrm{v},a,h}
 \in\mathbb{R}^{d\times d_\text{h}}$,
$d_\text{h}=d/n_\text{h}$, and $\mathrm{LN}(\cdot)$ denotes layer
normalization over the channel dimension. The output of the $h$th head for
slice $s$ is represented by
\begin{align}
    \mathbf{O}^{(\ell)}_{a,s,h}
    =
    \mathrm{softmax}\left(
        \frac{
        \mathbf{Q}^{(\ell)}_{a,s,h}
        \big(\mathbf{K}^{(\ell)}_{a,s,h}\big)^{\mathrm{T}}
        }{\sqrt{d_\text{h}}}
    \right)
    \mathbf{V}^{(\ell)}_{a,s,h},
    \label{eq:attn_head}
\end{align}
and the corresponding multi-head attention output reads
\begin{align}
    \mathrm{MHA}^{(\ell)}_{a}
    \left(\mathbf{H}^{(\ell-1)}_{a,s}\right)
    =
    \big[
        \mathbf{O}^{(\ell)}_{a,s,1},
        \ldots,
        \mathbf{O}^{(\ell)}_{a,s,n_\text{h}}
    \big]
    \mathbf{W}^{(\ell)}_{\mathrm{p},a},
    \label{eq:mha}
\end{align}
where
$\mathbf{W}^{(\ell)}_{\mathrm{p},a}\in\mathbb{R}^{d\times d}$.
The projection matrices are shared across all slices of the same axis within
a block. For $a=\mathrm{R}$, each attention matrix spans all $N_\text{c}$
range cells, allowing direct interaction between arbitrarily separated range
locations without imposing a locality constraint. For $a=\mathrm{D}$, the
same operation is applied over all $N_\text{sym}$ Doppler cells to model the
remaining Doppler-domain dependencies after DCF compensation.

For the $\ell$-th rejection block, the input
$\mathbf{H}^{(\ell-1)}\in
\mathbb{R}^{d\times N_\text{c}\times N_\text{sym}}$
is processed sequentially as follows.

\textit{Step 1) Range attention:}
For each Doppler index $\mu=0,\ldots,N_\text{sym}-1$, the range slice
$\mathbf{H}^{(\ell-1)}_{\mathrm{R},\mu}
\in\mathbb{R}^{N_\text{c}\times d}$
is updated through the range-axis multi-head attention as
\begin{align}
    \widetilde{\mathbf{H}}^{(\ell)}_{\mathrm{R},\mu}
    &=
    \mathbf{H}^{(\ell-1)}_{\mathrm{R},\mu}
    +
    \mathrm{MHA}^{(\ell)}_{\mathrm{R}}
    \left(
        \mathbf{H}^{(\ell-1)}_{\mathrm{R},\mu}
    \right).
    \label{eq:axial_r}
\end{align}
Collecting the updated range slices over all $\mu$ yields
$\widetilde{\mathbf{H}}^{(\ell)}
\in\mathbb{R}^{d\times N_\text{c}\times N_\text{sym}}$.

\textit{Step 2) Doppler attention:}
For each range index $m=0,\ldots,N_\text{c}-1$, the Doppler slice
$\widetilde{\mathbf{H}}^{(\ell)}_{\mathrm{D},m}
\in\mathbb{R}^{N_\text{sym}\times d}$
is then updated through the Doppler-axis multi-head attention as
\begin{align}
    \widehat{\mathbf{H}}^{(\ell)}_{\mathrm{D},m}
    &=
    \widetilde{\mathbf{H}}^{(\ell)}_{\mathrm{D},m}
    +
    \mathrm{MHA}^{(\ell)}_{\mathrm{D}}
    \left(
        \widetilde{\mathbf{H}}^{(\ell)}_{\mathrm{D},m}
    \right).
    \label{eq:axial_d}
\end{align}
Collecting the updated Doppler slices over all $m$ yields
$\widehat{\mathbf{H}}^{(\ell)}
\in\mathbb{R}^{d\times N_\text{c}\times N_\text{sym}}$.

\textit{Step 3) Pointwise MLP:}
Finally, the MLP is applied independently to the $d$-dimensional feature
vector at each range-Doppler cell:
\begin{align}
    \mathbf{H}^{(\ell)}
    &=
    \widehat{\mathbf{H}}^{(\ell)}
    +
    \mathrm{MLP}^{(\ell)}
    \left(
        \mathrm{LN}
        \left(
            \widehat{\mathbf{H}}^{(\ell)}
        \right)
    \right).
    \label{eq:axial_m}
\end{align}

This axial factorization avoids constructing a full
$N_\text{c}N_\text{sym}\times N_\text{c}N_\text{sym}$ attention matrix.
Instead, the range attention forms $N_\text{sym}$ matrices of size
$N_\text{c}\times N_\text{c}$, while the Doppler attention forms
$N_\text{c}$ matrices of size
$N_\text{sym}\times N_\text{sym}$. Consequently, the attention complexity
is reduced from
$O(N_\text{c}^2N_\text{sym}^2d)$
to
$O\left(
N_\text{c}N_\text{sym}(N_\text{c}+N_\text{sym})d
\right)$.

After the $L$th block, a convolution with the kernel $\boldsymbol{\mathcal{W}}_\text{o}\in\mathbb{R}^{d_\psi\times d\times3\times3}$ produces the ICI-mitigated data cube, which reads
\begin{align}\label{eq:head_out}
    \mathcal{F}_\psi = \phi\big(\boldsymbol{\mathcal{W}}_\text{o} * \mathbf{H}^{(L)}\big).
\end{align}

\subsection{Stage 3: Detection Head}
\label{subsec:imnet_detection}

The detection head maps $\mathcal{F}_\psi$ to a target-confidence map using
dilated convolutions followed by a pointwise projection and sigmoid activation:
\begin{align}
    \hat{\mathbf{M}}
    =
    \sigma\left(
        \mathcal{D}_\rho(\mathcal{F}_\psi)
    \right)
    \in [0,1]^{N_\text{c}\times N_\text{sym}},
    \label{eq:detection_head}
\end{align}
where $\mathcal{D}_\rho(\cdot)$ denotes the detection head with learnable
parameters $\rho$. Targets are declared by thresholding
$\hat{\mathbf{M}}$ at $\tau$, selected to satisfy the desired false-alarm
budget $P_{\mathrm{FA}}^\star$. For a detected peak
$(m^\star(i),\tilde{\mu}^\star(i))$, the range and velocity estimates are
\begin{align}
    \hat{R}_i
    &=
    \frac{m^\star(i)c_0}{2N_\text{c} \beta \Delta f},\\
    \hat{v}_i
    &=
    -\frac{\tilde{\mu}^\star(i)c_0\Delta f}
    {2\alpha N_\text{sym}f_\text{c}},
\end{align}
where $\tilde{\mu}^\star(i)$ denotes the signed Doppler-bin index.

\subsection{Training Objective}
\label{subsec:imnet_training}

IMNet is trained in a supervised manner over the learnable parameters
$\boldsymbol{\Theta}=\psi\cup\rho$, where $\psi$ collects all learnable parameters of
Stage~2, including the gating, embedding, positional encoding, axial-attention,
MLP, and output-projection parameters, and $\rho$ denotes the parameters of
the detection head.
Each training sample consists of the DCF-bank tensor
$\Bar{\boldsymbol{\mathcal{Y}}}^{\mathrm{DCF}(j)}$ and the corresponding
target parameters
$\{\bar\tau_i^{(j)},\bar f_{\mathrm{D},i}^{(j)}\}_{i\in\mathcal{P}}$.

The supervision is a binary target-presence map
$\mathbf{M}^{\star(j)}\in\{0,1\}^{N_\text{c}\times N_\text{sym}}$ defined as
\begin{align}
[\mathbf{M}^{\star(j)}]_{m,\mu}
=
\begin{cases}
1, & (m,\mu)=(m_i^{(j)},\mu_i^{(j)})
     \text{ for some } i\in\mathcal{P},\\
0, & \text{otherwise},
\end{cases}
\label{eq:ground_truth}
\end{align}
where $m_i^{(j)}$ and $\mu_i^{(j)}$ denote the range and Doppler bins
corresponding to $\bar\tau_i^{(j)}$ and $\bar f_{\mathrm{D},i}^{(j)}$,
respectively.

Since only a small fraction of the range-Doppler cells contain targets,
IMNet is trained using the focal loss \cite{Lin17},
\begin{align}
\mathcal{L}_\text{F}
=
-\sum_{\mu=0}^{N_\text{sym}-1}
\sum_{m=0}^{N_\text{c}-1}
(1-p_{m,\mu}^{(j)})^\gamma
\log p_{m,\mu}^{(j)},
\label{eq:focal_loss}
\end{align}
where $\gamma$ is the focusing parameter and
\begin{align}
p_{m,\mu}^{(j)}
=
\begin{cases}
[\hat{\mathbf{M}}^{(j)}]_{m,\mu},
& [\mathbf{M}^{\star(j)}]_{m,\mu}=1,\\
1-[\hat{\mathbf{M}}^{(j)}]_{m,\mu},
& \text{otherwise}.
\end{cases}
\end{align}
The parameters $\boldsymbol{\Theta}$ are optimized over mini-batches, while Stage~1
remains parameter free.


\subsection{Local Refinement}
\label{subsec:imnet_local_refinement}

The local-refinement version of IMNet, denoted IMNet-LR, improves the grid-limited IMNet detections to sub-cell resolution using the original signal model. The vectorized observation is first defined as
\begin{align}
    \mathbf{y}
    =
    \mathrm{vec}(\mathbf{Y})
    \in\mathbb{C}^{N_\text{c}N_\text{sym}},
    \label{eq:imnet_lr_observation}
\end{align}
where $\mathrm{vec}(\cdot)$ stacks the columns of its matrix argument.
For normalized Doppler $\nu$ and normalized delay $\tau$, the single-target
response is
\begin{align}
    \mathbf{A}(\nu,\tau)
    =
    \mathbf{D}_\text{I}(\nu)
    \mathbf{F}_{N_\text{c},\beta}^{H}
    \mathbf{D}_\text{R}^{*}(\beta\tau)
    \widetilde{\mathbf{S}}
    \mathbf{D}_\text{v}(\nu),
    \label{eq:imnet_lr_atom_matrix}
\end{align}
with the corresponding vectorized atom
\begin{align}
    \mathbf{b}(\nu,\tau)
    =
    \mathrm{vec}\left(\mathbf{A}(\nu,\tau)\right)
    \in\mathbb{C}^{N_\text{c}N_\text{sym}}.
    \label{eq:imnet_lr_atom}
\end{align}
Accordingly, the vectorized observation is modeled as
\begin{align}
    \mathbf{y}
    =
    \sum_{i\in\mathcal{P}}
    c_i\,\mathbf{b}(\nu_i,\tau_i)
    + \mathbf{z},
    \label{eq:imnet_lr_vector_model}
\end{align}
where $c_i$ collects the unknown complex path gain and angular response and $\mathbf{z} \sim \mathcal{CN}\left(\mathbf{0},\sigma^2\mathbf{I}_{N_\text{c}N_\text{sym}}\right)$.

For the $j$th detection, the coarse detection from IMNet can be represented by $\tau_j = m^\star(j) /N_\text{c}$ and $\nu_j = \tilde{\mu}^\star(j) /  (\alpha N_\text{sym})$.
To mitigate interference from stronger neighboring targets, the context
atoms associated with the $j$th seed are collected as
\begin{align}
    \mathbf{B}_j
    =
    \big[
        \mathbf{b}(\hat{\nu}_\ell,\hat{\tau}_\ell)
    \big]_{\ell\in\mathcal{C}_j},
    \qquad j\notin\mathcal{C}_j,
    \label{eq:imnet_lr_context}
\end{align}
where $\mathcal{C}_j=\{\ell\in\mathcal{P}\setminus\{j\}:
[\hat{\mathbf{M}}]_{m^\star(\ell),\mu^\star(\ell)}\geq T_\mathrm{I}\}$,
where $T_\mathrm{I}$ is the confidence threshold for selecting interfering
detections. Their subspace is projected out by
\begin{align}
    \mathbf{P}_j^\perp
    =
    \mathbf{I}_{N_\text{c}N_\text{sym}}
    -
    \mathbf{B}_j
    \left(
        \mathbf{B}_j^\mathrm{H}\mathbf{B}_j
    \right)^\dagger
    \mathbf{B}_j^\mathrm{H}.
    \label{eq:imnet_lr_projection}
\end{align}
For a candidate $(\nu,\tau)$ in the current local search grid, the
concentrated ML score is given by
\begin{align}
    \Lambda_j(\nu,\tau)
    =
    \frac{
        \left|
        \mathbf{b}^\mathrm{H}(\nu,\tau)
        \mathbf{P}_j^\perp\mathbf{y}
        \right|^2
    }{
        \mathbf{b}^\mathrm{H}(\nu,\tau)
        \mathbf{P}_j^\perp
        \mathbf{b}(\nu,\tau)
    }.
    \label{eq:imnet_lr_score}
\end{align}
Let $\Omega_j$ denote the fine-grid search, which is given by
\begin{align}
\Omega_j
=
\Bigg\{(\nu,\tau):
\left|\tau-\tau_j\right|
\leq \frac{1}{2N_\text{c}},
\;
\left|\nu-\nu_j\right|
\leq \frac{1}{2\alpha N_\text{sym}}
\Bigg\}.
\label{eq:imnet_lr_region}
\end{align}
The final sub-cell estimate is
\begin{align}
    (\hat{\nu}_j,\hat{\tau}_j)
    =
    \argmax_{(\nu,\tau)\in\Omega_j}
    \Lambda_j(\nu,\tau).
    \label{eq:imnet_lr_argmax}
\end{align}

\section{Transmit Beamforming Design}
\label{sec:bf_optimization}

In this section, transmit beamforming for the MU MIMO-SEFDM ISAC system is considered, where the objective is to maximize the multi-user communication sum-rate while maintaining a prescribed sensing beampattern gain. The corresponding constrained optimization problem is then formulated, followed by the development of an efficient iterative solution.

\subsection{Problem Formulation}
\label{subsec:bf_problem}

The BS designs $\{\mathbf{v}_k^{(n)}\}$ based on the frequency-domain channels 
$\{\mathbf{h}_k^{(n)}\}$. Since the SINR in \eqref{eq:sinr_def} depends on the 
beamformers of other subcarriers through the S-ICI term 
$J_k^{(n)}(\beta)$ in \eqref{def:J_kn}, the resulting sum-rate maximization is 
coupled across the entire band:
\begin{subequations}\label{eq:bf_robust}
\begin{alignat}{2}
    & \max_{\{\mathbf{v}_k^{(n)}\}} \quad
    && \sum_{n\in\mathcal{N}_\mathrm{c}} R_\mathrm{sum}^{(n)},
    \label{eq:bf_robust_a}\\
    & \text{s.t.} \quad
    && \mathrm{tr}\!\left(
        \sum_{k\in\mathcal{K}}
        \mathbf{v}_k^{(n)}
        \big(\mathbf{v}_k^{(n)}\big)^\mathrm{H}
    \right)
    \le P_\mathrm{tot}^{(n)},
    \quad \forall n\in\mathcal{N}_\mathrm{c},
    \label{eq:bf_robust_b}\\
    &&
    & \frac{1}{N_\mathrm{c}}
    \sum_{j=1}^{J}
    \sum_{k\in\mathcal{K}}
    \left|
        \mathbf{a}_\mathrm{T}^\mathrm{H}(\theta_j)
        \mathbf{v}_k^{(n)}
    \right|^2
    \ge P_\mathrm{T,req}^{(n)},
    \quad \forall n\in\mathcal{N}_\mathrm{c}.
    \label{eq:bf_robust_c}
\end{alignat}
\end{subequations}
where $R_\mathrm{sum}^{(n)}=\sum_{k\in\mathcal{K}} \log\left(1+\gamma_k^{(n)}\right)$.
The sensing constraint \eqref{eq:bf_robust_c} enforces a minimum aggregate
transmit beampattern gain over the $J$ focal directions on each subcarrier.
Thus, the constraints are separable across subcarriers, whereas the objective
remains coupled through the S-ICI.

\subsection{WMMSE Reformulation}
\label{subsec:bf_surrogate}

The WMMSE formulation \cite{Shi11_TSP}, which provides an equivalent representation of the sum-rate objective in \eqref{eq:bf_robust_a}, is adopted. Let user $k$ apply the receive scalar $u_k^{(n)}$ to its subcarrier-$n$ observation, and define $P_k^{(n)} = \sum_{\ell\in\mathcal{K}} \left| (\mathbf{h}_k^{(n)})^\mathrm{H} \mathbf{v}_\ell^{(n)} \right|^2.$
Then, \eqref{def:J_kn} can be written as
\begin{align}
    J_k^{(n)}(\beta)
    =
    \sum_{m\in\mathcal{N}_\mathrm{c}\setminus\{n\}}
    |\rho_{m,n}(\beta)|^2 P_k^{(m)} .
\end{align}
For unit-energy symbols, the mean-square error (MSE) of user $k$ on subcarrier $n$ is given by
\begin{align}\label{eq:mse_k}
    \mathrm{MSE}_k^{(n)}
    &=
    \left|
    1-(u_k^{(n)})^*
    (\mathbf{h}_k^{(n)})^\mathrm{H}
    \mathbf{v}_k^{(n)}
    \right|^2
    \nonumber\\
    & \hspace{-30pt}
    +
    |u_k^{(n)}|^2
    \Bigg(
    P_k^{(n)}
    -
    \left|
    (\mathbf{h}_k^{(n)})^\mathrm{H}
    \mathbf{v}_k^{(n)}
    \right|^2
    +
    J_k^{(n)}(\beta)
    +
    \sigma_k^2
    \Bigg).
\end{align}

Introducing the positive weights $\alpha_k^{(n)}>0$, the corresponding WMMSE problem is
\begin{align}\label{eq:wmmse_obj}
    \min_{\{\mathbf{v}_k^{(n)},u_k^{(n)},\alpha_k^{(n)}\}}
    \sum_{n\in\mathcal{N}_\mathrm{c}}
    \sum_{k\in\mathcal{K}}
    \left(
    \alpha_k^{(n)}\mathrm{MSE}_k^{(n)}
    -
    \log \alpha_k^{(n)}
    \right),
\end{align}
subject to \eqref{eq:bf_robust_b} and \eqref{eq:bf_robust_c}. The WMMSE formulation has the same stationary points in the precoders as the sum-rate problem \cite{Shi11_TSP}. For fixed precoders, the optimal receive scalar and MSE weight are
\begin{align}\label{eq:wmmse_updates}
    \hat{u}_k^{(n)}
    &=
    \frac{
    (\mathbf{h}_k^{(n)})^\mathrm{H}
    \mathbf{v}_k^{(n)}
    }{
    P_k^{(n)}
    +
    J_k^{(n)}(\beta)
    +
    \sigma_k^2
    },
    \\
    \hat{\alpha}_k^{(n)}
    &=
    \frac{1}{\mathrm{MSE}_k^{(n)}} ,
\end{align}
where $\hat{\alpha}_k^{(n)}=1+\gamma_k^{(n)}$ when
$u_k^{(n)}=\hat{u}_k^{(n)}$.

For fixed $\{u_k^{(n)},\alpha_k^{(n)}\}$, the only
cross-subcarrier dependence in the precoder update arises from the
SEFDM-induced leakage terms. These terms can be re-indexed as
\begin{align}\label{eq:eta_weight}
    &\sum_{n\in\mathcal{N}_\mathrm{c}}
    \sum_{k\in\mathcal{K}}
    \alpha_k^{(n)}
    |u_k^{(n)}|^2
    J_k^{(n)}(\beta)
    =
    \sum_{n\in\mathcal{N}_\mathrm{c}}
    \sum_{k\in\mathcal{K}}
    \eta_k^{(n)} P_k^{(n)},
\end{align}
where $\eta_k^{(n)}
    =
    \sum_{m\in\mathcal{N}_\mathrm{c}\setminus\{n\}}
    \alpha_k^{(m)}
    |u_k^{(m)}|^2
    |\rho_{n,m}(\beta)|^2.$
Thus, for fixed WMMSE auxiliary variables, the leakage generated by
subcarrier $n$ is collected into the scalar $\eta_k^{(n)}$, and the
precoder-dependent objective becomes separable across subcarriers
without approximation. Dropping terms independent of the precoders,
the reformulated optimization problem is given by
\begin{subequations}\label{eq:bf_robust_qp}
\begin{alignat}{2}
    & \min_{\mathbf{V}^{(n)}} \quad
    && \sum_{k\in\mathcal{K}}
    (\mathbf{v}_k^{(n)})^\mathrm{H}
    \mathbf{G}^{(n)}
    \mathbf{v}_k^{(n)}
    \nonumber\\
    & && \quad
    -2\sum_{k\in\mathcal{K}}
    \mathrm{Re}\!\left\{
    \alpha_k^{(n)}
    (u_k^{(n)})^*
    (\mathbf{h}_k^{(n)})^\mathrm{H}
    \mathbf{v}_k^{(n)}
    \right\},
    \label{eq:bf_robust_qp_obj}\\
    & \text{s.t.} \quad
    && \eqref{eq:bf_robust_b},\ \eqref{eq:bf_robust_c},
\end{alignat}
\end{subequations}
where
\begin{align}\label{eq:Gmat}
    \mathbf{G}^{(n)}
    =
    \sum_{p\in\mathcal{K}}
    \left(
    \alpha_p^{(n)}|u_p^{(n)}|^2
    +
    \eta_p^{(n)}
    \right)
    \mathbf{h}_p^{(n)}
    (\mathbf{h}_p^{(n)})^\mathrm{H}
    \succeq \mathbf{0}.
\end{align}

\subsection{Closed-Form Precoder}
\label{subsec:bf_closedform}

For fixed WMMSE auxiliary variables, the precoder subproblem \eqref{eq:bf_robust_qp} has a quadratic objective, while the sensing constraint is a reverse-convex quadratic constraint. A closed-form stationary precoder is derived from the Karush--Kuhn--Tucker (KKT) conditions.

The sensing requirement can be expressed through the aggregate matrix
$\mathbf{A}
    =
    \frac{1}{N_\mathrm{c}}
    \sum_{j=1}^{J}
    \mathbf{a}_\mathrm{T}(\theta_j)
    \mathbf{a}_\mathrm{T}^\mathrm{H}(\theta_j)
    \succeq \mathbf{0},
$
such that the left-hand side of \eqref{eq:bf_robust_c} is
$\sum_{k\in\mathcal{K}}
    (\mathbf{v}_k^{(n)})^\mathrm{H}
    \mathbf{A}
    \mathbf{v}_k^{(n)}.$
Then, by introducing the Lagrange multipliers
$\lambda^{(n)}\ge0$ and $\mu^{(n)}\ge0$ associated with
\eqref{eq:bf_robust_b} and \eqref{eq:bf_robust_c}, respectively,
the Lagrangian is given by
\begin{align}\label{eq:scalar_lagrangian}
    &\mathcal{L}
    =
    \sum_{k\in\mathcal{K}}
    (\mathbf{v}_k^{(n)})^\mathrm{H}
    \Big(
        \mathbf{G}^{(n)}
        +\lambda^{(n)}\mathbf{I}_{N_\mathrm{T}}
        -\mu^{(n)}\mathbf{A}
    \Big)
    \mathbf{v}_k^{(n)}
    \\
    &
    -2 \hspace{-2pt}\sum_{k\in\mathcal{K}}
    \mathrm{Re}\!\left\{
        \alpha_k^{(n)}
        (u_k^{(n)})^*
        (\mathbf{h}_k^{(n)})^\mathrm{H}
        \mathbf{v}_k^{(n)}
    \hspace{-2pt} \right\}
    \hspace{-2pt}-\hspace{-2pt}\lambda^{(n)} \hspace{-2pt} P_\mathrm{tot}^{(n)}
    \hspace{-2pt}+\hspace{-2pt}\mu^{(n)} \hspace{-2pt} P_\mathrm{T,req}^{(n)}. \nonumber
\end{align}

The first-order stationarity condition
$\partial\mathcal{L}/\partial(\mathbf{v}_k^{(n)})^*=\mathbf{0}$
gives
\begin{align}\label{eq:v_lagrange}
    \mathbf{v}_k^{(n)}
    =
    \alpha_k^{(n)}u_k^{(n)}
    \Big(
        \mathbf{G}^{(n)}
        +\lambda^{(n)}\mathbf{I}_{N_\mathrm{T}}
        -\mu^{(n)}\mathbf{A}
    \Big)^{-1}
    \mathbf{h}_k^{(n)} .
\end{align}
The multipliers $\lambda^{(n)}$ and $\mu^{(n)}$ are then determined
from the corresponding complementary-slackness conditions.

\subsection{Solution for the Lagrange Multipliers}
\label{subsec:bf_closedform_xi}

The multipliers $\lambda^{(n)}$ and $\mu^{(n)}$ are optimized alternately, and this procedure is repeated iteratively until convergence.

\paragraph{Solution for $\lambda^{(n)}$}
With the fixed $\mu^{(n)}$, $\mathbf{v}_k^{(n)}(\lambda^{(n)},\mu^{(n)})$ denotes the right-hand side of \eqref{eq:v_lagrange}. If $\mathbf{G}^{(n)}-\mu^{(n)}\mathbf{A}\succ\mathbf{0}$ and $\sum_{k\in\mathcal{K}}\|\mathbf{v}_k^{(n)}(0,\mu^{(n)})\|_2^2 \le P_\text{tot}^{(n)}$, $\hat\lambda^{(n)}=0$ is obtained. Otherwise, the constraint \eqref{eq:bf_robust_b} must hold with equality. Let $\mathbf{E}\boldsymbol{\Lambda}\mathbf{E}^\text{H}$ represent the eigenvalue decomposition of $\mathbf{G}^{(n)}-\mu^{(n)}\mathbf{A}$, so that the matrix inverted in \eqref{eq:v_lagrange} becomes $\mathbf{E}(\boldsymbol{\Lambda}+\lambda^{(n)}\mathbf{I}_{N_\text{T}})\mathbf{E}^\text{H}$. 
To ensure that the Lagrangian is bounded below and the inverse in \eqref{eq:v_lagrange} is well defined, $\lambda^{(n)}$ is restricted to
\begin{align}\label{eq:lambda_range}
    \lambda^{(n)} > \max\left\{0,-\lambda_{\min}\big(\mathbf{G}^{(n)}-\mu^{(n)}\mathbf{A}\big)\right\}.
\end{align}
Then, the following equation should be satisfied:
\begin{align}\label{eq:eig_power}
    \sum_{k\in\mathcal{K}}\mathrm{Tr}\Big(\big(\boldsymbol{\Lambda}+\lambda^{(n)}\mathbf{I}_{N_\text{T}}\big)^{-2}\boldsymbol{\Upsilon}_k\Big) = P_\text{tot}^{(n)},
\end{align}
where $\boldsymbol{\Upsilon}_k = \big(\alpha_k^{(n)}\big)^2\big|u_k^{(n)}\big|^2\mathbf{E}^\text{H}\mathbf{h}_k^{(n)}(\mathbf{h}_k^{(n)})^\text{H}\mathbf{E}$. Then, \eqref{eq:eig_power} is equivalent to
\begin{align}\label{eq:opt_lambda}
    \sum_{k\in\mathcal{K}}\sum_{i=1}^{N_\text{T}}\frac{[\boldsymbol{\Upsilon}_k]_{ii}}{\big([\boldsymbol{\Lambda}]_{ii}+\lambda^{(n)}\big)^2} = P_\text{tot}^{(n)}.
\end{align}
Over the admissible interval in \eqref{eq:lambda_range}, the left-hand side of \eqref{eq:opt_lambda} is monotonically decreasing in $\lambda^{(n)}$. Hence, $\hat\lambda^{(n)}$ can be readily obtained via a one-dimensional bisection method.

\paragraph{Solution for $\mu^{(n)}$}
With the fixed $\lambda^{(n)}$, if the sensing constraint \eqref{eq:bf_robust_c} is satisfied at $\mu^{(n)}=0$, it holds that $\hat{\mu}^{(n)}=0$. Otherwise, the sensing constraint is active and must hold with equality. To express this equality explicitly in terms of $\mu^{(n)}$, $\mathbf{M}^{(n)} = \mathbf{G}^{(n)}+\lambda^{(n)}\mathbf{I}_{N_\text{T}}$ is defined, so that \eqref{eq:v_lagrange} can be written as $\mathbf{v}_k^{(n)}=\alpha_k^{(n)}u_k^{(n)}(\mathbf{M}^{(n)}-\mu^{(n)}\mathbf{A})^{-1}\mathbf{h}_k^{(n)}$. Since the direct substitution of this expression into the sensing constraint involves the inverse of $\mathbf{M}^{(n)}-\mu^{(n)}\mathbf{A}$, its $\mu^{(n)}$-dependent part is diagonalized through the eigenvalue decomposition
\begin{align}\label{eq:eig_whitened}
    (\mathbf{M}^{(n)})^{-1/2}\mathbf{A}(\mathbf{M}^{(n)})^{-1/2}
    = \mathbf{Q}\boldsymbol{\Gamma}\mathbf{Q}^\text{H},
\end{align}
where $\mathbf{Q}$ is unitary and $\boldsymbol{\Gamma}=\mathrm{diag}(\gamma_1,\ldots,\gamma_{N_\text{T}})$ contains the nonnegative eigenvalues. Accordingly,
$(\mathbf{M}^{(n)}-\mu^{(n)}\mathbf{A})^{-1}
    =
    (\mathbf{M}^{(n)})^{-1/2}
    \mathbf{Q}
    (\mathbf{I}_{N_\text{T}}-\mu^{(n)}\boldsymbol{\Gamma})^{-1}
    \mathbf{Q}^\text{H}
    (\mathbf{M}^{(n)})^{-1/2}.$
Writing $\mathbf{z}_k=\alpha_k^{(n)}u_k^{(n)}\mathbf{Q}^\text{H}(\mathbf{M}^{(n)})^{-1/2}\mathbf{h}_k^{(n)}$, the sensing-gain equality reduces to the scalar form
\begin{align}\label{eq:gain_scalar}
    \sum_{k\in\mathcal{K}}(\mathbf{v}_k^{(n)})^\text{H}\mathbf{A}\mathbf{v}_k^{(n)}
    =
    \sum_{i=1}^{N_\text{T}}
    \frac{\gamma_i \sum_{k\in\mathcal{K}}|[\mathbf{z}_k]_i|^2}{(1-\mu^{(n)}\gamma_i)^2}.
\end{align}
Denoting the largest eigenvalue by $\gamma_\text{max} = \max_i\gamma_i$, the Lagrangian \eqref{eq:scalar_lagrangian} is bounded below only while $\mathbf{I}_{N_\text{T}}-\mu^{(n)}\boldsymbol{\Gamma}\succ\mathbf{0}$, which restricts $\mu^{(n)}$ to
\begin{align}\label{eq:mu_range}
    0 \le \mu^{(n)} < \mu_\text{max}^{(n)} = \frac{1}{\gamma_\text{max}} .
\end{align}
If $\sum_{k\in\mathcal{K}}(\mathbf{v}_k^{(n)})^\text{H}\mathbf{A}\mathbf{v}_k^{(n)} \ge P_\text{T,req}^{(n)}$ already holds at $\mu^{(n)}=0$, it follows that $\hat\mu^{(n)}=0$. Otherwise, the constraint \eqref{eq:bf_robust_c} must hold with equality, so that \eqref{eq:gain_scalar} gives
\begin{align}\label{eq:mu_eq}
    \sum_{i=1}^{N_\text{T}}\frac{\gamma_i \sum_{k\in\mathcal{K}}|[\mathbf{z}_k]_i|^2}{(1-\mu^{(n)}\gamma_i)^2} = P_\text{T,req}^{(n)} .
\end{align}
The left-hand side of \eqref{eq:mu_eq} is monotonically increasing in $\mu^{(n)}$ over \eqref{eq:mu_range}. Hence, $\hat\mu^{(n)}$ can be readily obtained via a one-dimensional bisection method.

\subsection{Overall Algorithm}
\label{subsec:bf_algorithm}

Algorithm~\ref{alg:isac_bf} summarizes the proposed precoder computation. The dominant computational cost arises from the $N_\text{T}\times N_\text{T}$ eigenvalue decompositions required in the multiplier updates, resulting in $\mathcal{O}(N_\text{c}N_\text{T}^3)$ complexity per inner iteration. The equations for $\lambda^{(n)}$ and $\mu^{(n)}$ are solved efficiently by one-dimensional bisection using \eqref{eq:opt_lambda} and \eqref{eq:mu_eq}, respectively. The procedure is repeated until the achieved sum rate converges.

\begin{algorithm}[t]
    \caption{WMMSE Beamforming for MU MIMO-SEFDM ISAC}
    \label{alg:isac_bf}
    \begin{algorithmic}[1]
        \State \textbf{Input:} Channels $\{\mathbf{h}_k^{(n)}\}$, focal angles $\{\theta_j\}$, per-subcarrier budgets $\{P_\text{tot}^{(n)}\}$, sensing thresholds $\{P_\text{T,req}^{(n)}\}$, compression factor $\beta$.
        \State \textbf{Output:} Optimized precoders $\{\mathbf{v}_k^{(n)}\}$.
        \vspace{3pt}
        \State Initialize $\mathbf{V}$ and compute $\mathbf{A}$.
        \Repeat
            \State $u_k^{(n)},\,\alpha_k^{(n)} \gets$ \eqref{eq:wmmse_updates} for all $k,n$
            \State Compute $\eta_k^{(n)}=\sum_{m\ne n}\alpha_k^{(m)}|u_k^{(m)}|^2|\rho_{n,m}(\beta)|^2$ for all $k,n$, then $\mathbf{G}^{(n)} \gets$ \eqref{eq:Gmat}
            \ForAll{$n \in \mathcal{N}_\text{c}$ \textbf{in parallel}}
                \State $\mu^{(n)} \gets 0$
                \Repeat
                    \State \multiline{Compute $\mathbf{E}\boldsymbol{\Lambda}\mathbf{E}^\text{H}$ of $\mathbf{G}^{(n)}-\mu^{(n)}\mathbf{A}$ and $\boldsymbol{\Upsilon}_k$.}
                    \State \multiline{$\lambda^{(n)} \gets 0$ if $\mathbf{G}^{(n)}-\mu^{(n)}\mathbf{A}\succ\mathbf{0}$ and $\sum_{k\in\mathcal{K}}\|\mathbf{v}_k^{(n)}(0,\mu^{(n)})\|_2^2 \le P_\text{tot}^{(n)}$; otherwise, obtain $\lambda^{(n)}$ from \eqref{eq:opt_lambda} by bisection over its admissible interval.}
                    \State \multiline{Form $\mathbf{M}^{(n)}=\mathbf{G}^{(n)}+\lambda^{(n)}\mathbf{I}_{N_\text{T}}$ and compute $\mathbf{Q}\boldsymbol{\Gamma}\mathbf{Q}^\text{H}$ from \eqref{eq:eig_whitened}.}
                    \State \multiline{$\mu^{(n)} \gets 0$ if \eqref{eq:bf_robust_c} holds at $\mu^{(n)}=0$; otherwise, obtain $\mu^{(n)}$ from \eqref{eq:mu_eq} by bisection over \eqref{eq:mu_range}.}
                \Until{$\lambda^{(n)}$ and $\mu^{(n)}$ converge}
                \State Compute $\mathbf{v}_k^{(n)}$ from \eqref{eq:v_lagrange} for all $k \in \mathcal{K}$.
            \EndFor
        \Until{$\sum_n R_\text{sum}^{(n)}$ converges}
    \end{algorithmic}
\end{algorithm}

\section{Simulation Results}
\label{sec:simulation}
\paragraph{System parameters} The simulation setup follows~\cite{DCFNet_TWC2026} and extends it to SEFDM compression. Unless otherwise stated, the carrier frequency, bandwidth, subcarrier count, and symbol count are set to $f_\text{c}=60$~GHz, $B=50$~MHz, $N_\text{c}=2048$, and $N_\text{sym}=64$, respectively, which give a subcarrier spacing of $24.41$~kHz and FFT-limited resolutions of $\Delta R=3$~m and $\Delta v_\text{rel}=0.76$~m/s. The BS uses $N_\mathrm{T}=N_\mathrm{R}=20$ antennas. The compression factor is $\beta=0.6$. Each scene places $K=4$ users at a range of $40$~m moving at walking speed and $J=10$ targets at the focal angle $\theta_0=0$ with ranges drawn from $50$ to $400$~m. Transmit power $P_\mathrm{BS}=30$~dBm, sensing beampattern gain $P_\mathrm{T,req}=40$~dB, and noise variance $\sigma_k^2=-80$~dBm are used, and the channels follow 3GPP TR 38.901 \cite{3gpp38901}. All simulations run on an AMD Ryzen\textsuperscript{\texttrademark} Threadripper 7970X processor and an NVIDIA RTX PRO 6000 graphics processing unit (GPU).

\paragraph{Model parameters} IMNet uses a $9$-window DCF bank, a squeeze-and-excitation block with $r=2$, and a rejection head of $L=4$ axial blocks with width $96$ and $4$ heads returning $d_\psi=128$ channels, followed by a detection head of four $3\times3$ convolutions with dilations $1$, $2$, $4$, and $8$, which covers $31\times31$ range-Doppler cells. Training uses the focal loss with $\gamma=2$, $10{,}000$ training and $1{,}000$ test scenes, $500$ epochs, batch size $16$, and Adam with learning rate $10^{-4}$.

\begin{figure}[t]
\centering
\includegraphics[width=\columnwidth]{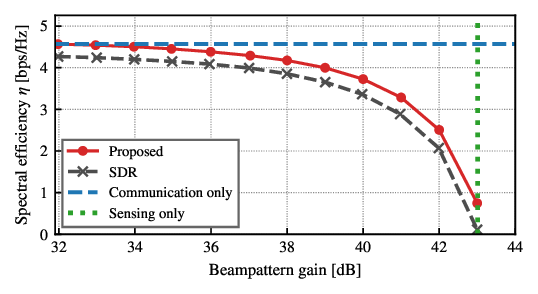}
\vspace{-10pt}
\caption{Spectral-efficiency/beampattern-gain trade-off for the proposed and baseline methods.}
\label{fig:SR_BG}
\end{figure}

\begin{figure}[t]
\centering
\includegraphics[width=\columnwidth]{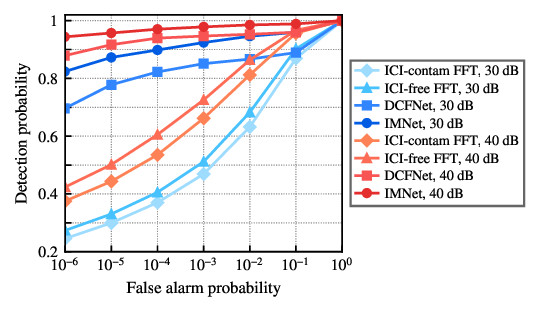}
\vspace{-20pt}
\caption{ROC curves for IMNet and baseline methods.}
\label{fig:RoC}
\end{figure}

\begin{figure*}[t]
\centering
\includegraphics[width=\textwidth]{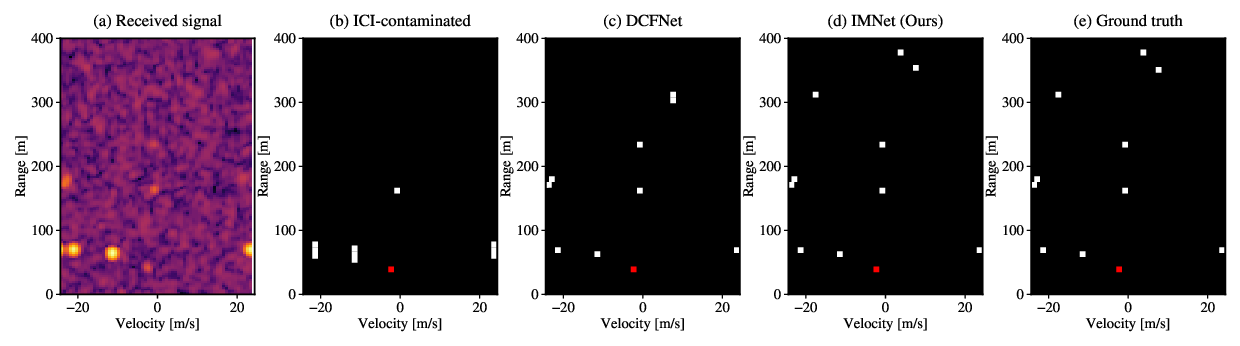}
\vspace{-10pt}
\caption{Comparison of range--velocity detection maps for IMNet and the baseline methods. For each detector, the $10$ strongest local peaks, excluding those corresponding to communication users, are declared as target candidates and shown in white, while the communication-user locations are indicated in red.}
\label{fig:RD_maps}
\end{figure*}

\begin{figure*}[t]
\centering
\includegraphics[width=\textwidth]{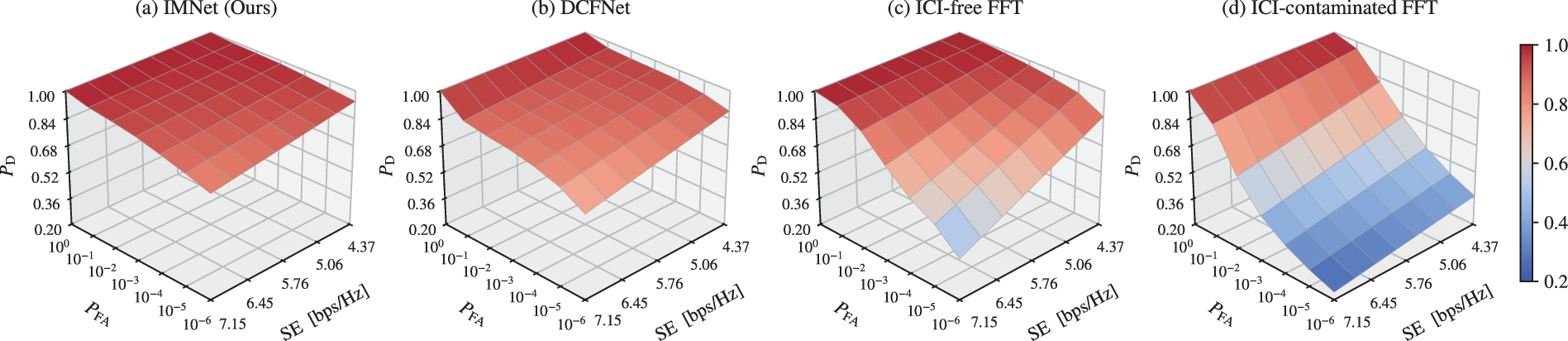}
\caption{Detection probability over false-alarm probability and spectral efficiency.}
\label{fig:surf_sr_dp_fp}
\end{figure*}

\subsection{Analysis on the Proposed Beamformers}

The proposed transmit beamformer is compared with the following three baselines.
\begin{itemize}[leftmargin=*]
    \item \textbf{Semidefinite relaxation (SDR)}: The rank-one constraint on the transmit covariance matrix is relaxed, and beamforming vectors are recovered from the optimized covariance matrix.
    \item \textbf{Sensing only}: The transmit beamformer is designed solely to maximize the sensing beampattern gain under the power constraint.
    \item \textbf{Communication only}: The sum-rate maximization problem in \eqref{eq:bf_robust} is solved without the sensing constraint in \eqref{eq:bf_robust_c}.
\end{itemize}

Fig.~\ref{fig:SR_BG} illustrates the trade-off between communication spectral efficiency and sensing beampattern gain as the minimum sensing requirement is varied. The proposed beamformer smoothly spans the operating region between the communication-only and sensing-only, with higher sensing requirements leading to increased beampattern gain at the expense of spectral efficiency. Moreover, the proposed method consistently achieves higher spectral efficiency than SDR for the same beampattern gain. This performance gap arises from the rank relaxation in SDR, where the optimized covariance matrix generally requires rank-one recovery to obtain feasible beamforming vectors, resulting in a loss relative to the proposed direct beamformer design. These results demonstrate that the proposed method provides a more favorable communication--sensing trade-off than the SDR baseline.

\subsection{Detection Performance of IMNet Compared to the Baseline Methods}
\label{subsec:sim_detection}

The following baselines are considered:
\begin{itemize}[leftmargin=*]
    \item \textbf{ICI-contaminated FFT}: 2D FFT applied directly to the received signals.
    \item \textbf{ICI-free FFT}: 2D FFT applied to ICI-free received signals, i.e., $\Bar{\mathbf{Y}}_{\text{D-ICI},i} = \Bar{\mathbf{Y}}_{\text{S-ICI},i} = 0, \forall i$.
    \item \textbf{DCFNet}: the $9$-window DCF bank, U-Net-based ICI-rejection module, and detection head from \cite{DCFNet_TWC2026}.
\end{itemize}
For the baseline methods without a dedicated target-detection rule, a conventional constant false alarm rate (CFAR) detector is applied to their sensing outputs for a fair comparison. The detection performance is characterized by the detection probability and false alarm probability, defined as
\begin{itemize}[leftmargin=*]
    \item \textbf{Detection probability ($P_\text{D}$)}: the fraction of actual targets that are correctly detected.
    \item \textbf{False alarm probability ($P_\text{FA}$)}: the number of false detections, i.e., detections not associated with any actual target, normalized by the total number of sensing-map cells.
\end{itemize}
By varying the CFAR threshold, different $(P_\text{FA},P_\text{D})$ operating points are obtained to construct the receiver operating characteristic (ROC) curves.

In Fig.~\ref{fig:RoC}, the ROC curves are plotted at $\beta=0.6$ for $P_\mathrm{T,req}\in\{30,40\}$~dB. The ICI-contaminated FFT baseline degrades sharply in the low-$P_\mathrm{FA}$ regime because the structured sidelobes and SEFDM-induced leakage raise the interference floor. The ICI-free FFT scheme substantially improves the ROC performance by removing the structured interference before detection. DCFNet and IMNet further outperform the ICI-free scheme, indicating the advantage of learned detection over conventional CFAR even in the absence of structured ICI. Between the two learning-based methods, IMNet achieves the highest detection probability in the low-$P_\mathrm{FA}$ regime. This gain is attributed to its range-axis attention, which is better suited than the local U-Net processing of DCFNet to suppress the S-ICI along the range dimension.

To provide qualitative insight into the detection behavior, Fig.~\ref{fig:RD_maps} visualizes a dense range--velocity scene with $J=10$ targets. The ICI-contaminated map exhibits strong sidelobes and numerous interference-induced false alarms, whereas DCFNet partially suppresses the interference through its DCF bank and U-Net-based rejection module. In contrast, IMNet more effectively suppresses both D-ICI and S-ICI, revealing weak target responses that were previously buried under the noise and interference floor. Consequently, even targets located far from the detector become detectable, and the resulting detections align more closely with the ground-truth range--velocity cells.

The joint sensing--communication trade-off is further examined in Fig.~\ref{fig:surf_sr_dp_fp}, which shows $P_\mathrm{D}$ as a function of $P_\mathrm{FA}$ and spectral efficiency $\eta$. As the false-alarm requirement becomes more stringent and the operating point shifts toward higher spectral efficiency, the detection probability generally decreases because less favorable sensing conditions make weak target responses more difficult to distinguish. This degradation is particularly pronounced for the ICI-contaminated FFT detector, whose $P_\mathrm{D}$ drops rapidly in the low-$P_\mathrm{FA}$ region. Removing the structured interference substantially improves the ICI-free FFT baseline, but its CFAR detector still exhibits a marked loss under stringent false-alarm constraints. The learning-based detectors are considerably more robust over the same operating region, with DCFNet maintaining a high $P_\mathrm{D}$ over a broader range of $\eta$. IMNet provides the most stable surface and preserves a high detection probability even in the challenging low-$P_\mathrm{FA}$ and high-$\eta$ regime, demonstrating that its improved ICI suppression enables sensing performance to be retained at higher communication spectral efficiency.

\subsection{Estimation Performance of IMNet Compared to The Baseline Methods}
\label{subsec:estimation}

\begin{figure}[t]
\centering
\includegraphics[width=\columnwidth]{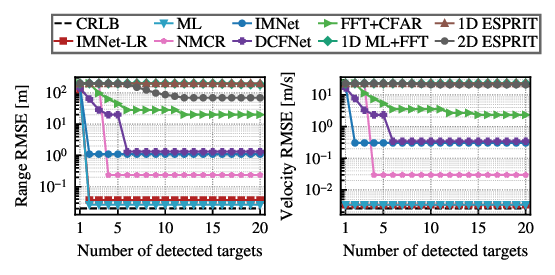}
\vspace{-20pt}
\caption{Weak-target range and velocity RMSE versus the number of detected targets. }
\label{fig:range_RMSE_per_target}
\end{figure}

In this subsection, the range and velocity root-mean-square error (RMSE) of IMNet, IMNet-LR, DCFNet, conventional ML, NMCR \cite{Mirabella26_TWC}, ESPRIT, and FFT-based methods is evaluated under the localization setting. A challenging two-target scenario is considered in which both targets have the same radial velocity of $23.19~\mathrm{m/s}$, and hence occupy the same Doppler bin. The strong target is located at $50$~m with a signal-to-noise ratio (SNR) of $20$~dB, whereas the weak target is located at $200$~m. Two complementary experiments are conducted: the weak-target SNR is fixed at $-10$~dB while the number of declared candidates $C$ is varied, and $C$ is fixed at $20$ while the weak-target SNR is swept from $-15$ to $10$~dB. Because both targets experience a large Doppler shift, the strong reflector generates substantial D-ICI that spreads across neighboring subcarriers and raises the interference floor around the weak target. In addition, the non-orthogonal SEFDM waveform introduces S-ICI across the range dimension, further obscuring the weak echo and making its localization particularly challenging.

In Fig.~\ref{fig:range_RMSE_per_target}, the weak-target SNR is fixed at $-10$~dB, while the number of declared candidates $C$ is varied from $1$ to $20$. For the score-map-based methods, candidate cells are ranked by confidence and the first $C$ local peaks are retained, whereas ML and 1D ML+FFT rank the refined candidates according to their final likelihood scores. With $C=1$, all methods select the strong target, resulting in a weak-target range error of approximately $150$~m, i.e., the separation between the two targets. As $C$ increases to $2$, IMNet-LR successfully identifies the weak target and reduces its range RMSE to $0.038$~m. This sub-cell accuracy is enabled by the local refinement in Sec.~\ref{subsec:imnet_local_refinement}, which uses the IMNet detection as a coarse seed and performs a fine-grid matched-filter ML search within its local neighborhood while projecting out the nuisance contribution of the strong target. Exhaustive ML exhibits a similar behavior and achieves a range RMSE of $0.027$~m, but at the cost of prohibitively high computational complexity. In contrast, NMCR requires $C=4$ before the weak target is included among its candidates because the dominant target and its sidelobes occupy the highest-ranked peaks, after which its range RMSE settles at approximately $0.237$~m. Without local refinement, the learning-based methods remain limited by the range--velocity grid resolution: IMNet reaches a range RMSE of approximately $1.10$~m from $C=2$ onward, whereas DCFNet requires $C=6$ to recover the weak target and then settles at about $1.32$~m. FFT+CFAR remains strongly affected by the structured interference and still exhibits a range RMSE of $20.0$~m at $C=20$, while the two ESPRIT variants and 1D ML+FFT fail to recover the weak target within the considered declaration budget.

\begin{figure}[t]
\centering
\includegraphics[width=\columnwidth]{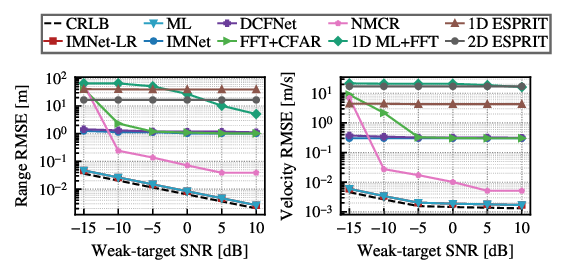}
\vspace{-20pt}
\caption{Weak-target range and velocity RMSE performance of the detection methods with respect to the SNR of the weak target. }
\label{fig:strong_weak_rmse_sweep}
\end{figure}

Fig.~\ref{fig:strong_weak_rmse_sweep} further evaluates the weak-target estimation accuracy by sweeping the weak-target SNR from $-15$ to $10$~dB. All methods are allowed up to $C=20$ declarations, so the comparison primarily reflects localization accuracy once the weak target has been identified. Range RMSE of IMNet-LR closely follows the Cram\'er--Rao lower bound (CRLB) over the entire SNR range, decreasing from $0.047$~m at $-15$~dB to $0.0027$~m at $10$~dB, with exhaustive ML achieving comparable accuracy. NMCR achieves relatively low estimation error, but remains inferior to IMNet-LR since it relies on the assumption that D-ICI is negligible. In the considered high-Doppler scenario, this assumption is violated, and the resulting D-ICI from the strong target masks the weak echo and limits the estimation accuracy of NMCR. The unrefined score-map methods remain limited by the range--velocity grid resolution, with RMSEs of approximately $1.1$--$1.4$~m across the SNR sweep, whereas FFT+CFAR is substantially more sensitive to the low-SNR regime and reaches $41.3$~m at $-15$~dB.

\begin{figure}[t]
\centering
\includegraphics[width=0.95\columnwidth]{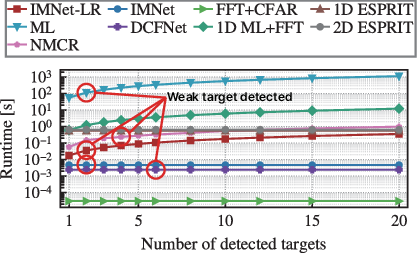}
\caption{Execution time for various numbers of detected targets. The red mark represents when the weak target is detected.}
\label{fig:range_RMSE_execution_time}
\end{figure}

The execution times of the considered estimators are compared in Fig.~\ref{fig:range_RMSE_execution_time}, where the red markers indicate the smallest number of declarations $C$ at which each method detects the weak target. Exhaustive ML detects the weak target at $C=2$, but already requires on the order of $10^{2}$~s because of its exhaustive fine-grid likelihood search over the entire range--Doppler domain. IMNet-LR also detects the weak target at $C=2$, while requiring only a few tens of milliseconds by restricting the ML refinement to a local neighborhood around the IMNet detection. Plain IMNet achieves detection at the same declaration budget with only a few milliseconds of execution time, although its estimation accuracy remains limited by the range--velocity grid resolution. NMCR requires $C=4$ before recovering the weak target and incurs a sub-second execution time, whereas DCFNet requires a larger declaration budget despite its millisecond-level complexity. These results highlight the main advantage of IMNet-LR: it preserves the early weak-target detection of IMNet and attains the near-ML estimation accuracy observed in Fig.~\ref{fig:strong_weak_rmse_sweep}, while reducing the execution time by more than three orders of magnitude relative to exhaustive ML.

\section{Conclusion}
This paper developed a MU MIMO-SEFDM ISAC framework that improves communication spectral efficiency while mitigating both D-ICI and S-ICI. A WMMSE-based beamformer with separable per-subcarrier precoder updates was derived, and IMNet, which combines a DCF bank with axial attention for effective ICI suppression and target detection, was proposed. IMNet-LR further refines the detected targets through local matched-filter ML with nuisance projection to achieve sub-cell range and velocity estimation. Simulations show that the proposed framework provides robust detection under severe interference, while IMNet-LR attains near-ML accuracy with more than three orders of magnitude lower execution time.

\bibliographystyle{IEEEtran}
\bibliography{bibtex}

\end{document}